\documentclass[oldversion]{aa}  
\usepackage{graphicx}
\usepackage{txfonts}
\usepackage{longtable}
\usepackage{natbib}
\usepackage{ulem}
\bibpunct{(}{)}{;}{a}{}{,}
\begin{document}
\title{The Environmental Dependence of Galaxy Properties at $z=2$}
\subtitle{}
\titlerunning{The Environmental Dependence of Galaxy Properties at $z=2$}
\authorrunning{Tanaka et al.}

\author{%M. Tanaka\inst{1} et al.
M. Tanaka\inst{1,2}, C. De Breuck\inst{1}, B. Venemans\inst{1}, and J. Kurk\inst{3}
}

\offprints{M. Tanaka}

\institute{European Southern Observatory, Karl-Schwarzschild-Str. 2
	D-85748 Garching bei M\"{u}nchen, Germany
	\email{mtanaka@eso.org}
	\and
	Institute for the Physics and Mathematics of the Universe, The University of Tokyo, 5-1-5 Kashiwanoha, Kashiwa-shi, Chiba 277-8583, Japan
 	\and
 	Max-Planck-Institut f\"{u}r extraterrestrische Physik, Giessenbachstra\ss e,
 	D-85748 Garching bei M\"{u}nchen, Germany
}

\date{Received; accepted }

\abstract{
We report on the environmental dependence of galaxy properties at $z=2.15$.
We construct multi-band photometric data sets in the (proto-)cluster PKS1138-26
field and in the GOODS field.  We then fit spectral energy distributions of
the galaxies with model templates generated with the latest stellar population
synthesis code and derive physical properties of galaxies from the fits.
To quantify the environmental dependence of galaxy properties, a special
care is taken of systematic errors --
we use data sets that have almost the same wavelength samplings,
use the same code to fit SEDs with the same set of templates, and compare
{\it relative} differences between the two samples.
We find that the PKS1138 galaxies have similar ages, shorter star formation
time scales, lower star formation rates, and weaker dust extinction
compared to the GOODS galaxies at $z\sim2$.
This trend is similar to that observed locally, suggesting that
the environmental dependence of galaxy properties is
already partly in place as early as $z=2.15$.
We show that the PKS1138 galaxies assemble
the bulk of their masses $\sim1$ Gyr earlier than field galaxies, i.e.,
the galaxy formation depends on environment.
Galaxy mergers should frequently occur during the first collapse of clusters
and they might play an important role in driving the observed environmental
dependence of galaxy properties at $z=2.15$.
}{}{}{}{}
% 5 {} token are mandatory
 
%\abstract{}


\maketitle

%-------------------------------------------------------
\section{Introduction}

The formation and evolution of galaxies in the Universe are
dependent on environment in which galaxies live.
In the local Universe, red early-type galaxies are
the dominant population in rich galaxy clusters,
while blue late-type galaxies are the dominant population in the low-density field.
Not only galaxy properties, but the formation epoch of
galaxies also depends on environment in the sense that
cluster galaxies form earlier than field galaxies
(e.g., \citealt{kuntschner02,gebhardt03,thomas05}).
Earlier studies concentrated on nearby galaxies, but with the recent advent of
large telescopes, environment studies at $z\sim1$ became possible.
Interestingly, the environmental dependence of galaxy properties
observed at $z\sim1$ is already strong;
clusters at $z\sim1$ are dominated by red early-type galaxies
(e.g., \citealt{blakeslee03,nakata05,postman05,lidman08,mei09},
but see also \citealt{cucciati06}).
Also, the formation epoch of cluster galaxies measured at $z\sim1$
is consistent with that observed locally \citep{gobat08} .
Although there is a clear sign of galaxy evolution between
$z=1$ and 0 (e.g., \citealt{elbaz07,cooper08}),
one has to observe galaxies at even higher redshifts to fully quantify
the environmental dependence of galaxy formation and evolution.

Because of observational difficulties, only a few high redshift
clusters are known so far, the highest redshift cluster being
at $z=1.45$ \citep{stanford06}.
Higher redshift galaxies appear fainter and their rest-frame optical
light migrates to the near-IR, where the sky background is brighter and
it is challenging to observe faint galaxies.
Furthermore, high redshift clusters are poor clusters 
as they are still fast growing according to the dark matter
halo growth models \citep{press74,springel05}.
Such poor clusters are difficult to locate due to their weak
density contrasts to the general field.
There are a number of ways to find high redshift clusters, but
one of the proven techniques is to look around high redshift
radio galaxies \citep{miley08}.
While not all the radio galaxies are in over-density regions,
many of them host clear over-densities of galaxies around them \citep{venemans07}
and they are called proto-clusters.
Although over-densities of red massive galaxies are not necessarily confirmed around them,
they likely virialize and evolve to clusters at lower redshifts.
Among the several proto-clusters reported so far, PKS1138-26 at $z=2.15$
is one of the most promising proto-clusters for its clear over-density of
spectroscopically confirmed galaxies by previous studies \citep{miley08}.

Early studies of the the PKS1138-26 radio galaxy
at $z=2.15$ were performed by \citet{pentericci97,pentericci98}, who
reported a clumpy morphology of the radio galaxy.
Followed by these initial observations,
\citet{kurk00} first reported an over-density of star forming galaxies
around the radio galaxy.
This region was then followed up by several authors.
\citet{pentericci00} performed spectroscopic follow-up observations of
Lyman $\alpha$ emitters reported in \citet{kurk00} and confirmed 14
galaxies close to the radio galaxy redshift.
\citet{kurk04a,kurk04b} carried out further imaging and
spectroscopic observations of the field targeting H$\alpha$ emitters
and confirmed another 10 galaxies at the cluster redshift.
An X-ray observation has also been performed \citep{pentericci02}
and follow-up spectroscopy confirmed  at least 5 X-ray sources
at the cluster redshift \citep{croft05}.
By now, there are more than 20 objects confirmed at the cluster redshift.
Detailed analyses of the radio galaxy and the surrounding region with
the superb resolution imaging with HST have also been performed
\citep{miley06,zirm08,hatch08,hatch09}, which added further lines
of evidence for the forming (proto-)cluster.
In fact, \citet{zirm08} reported on the forming cluster red sequence.
Recently, \citet{doherty09} carried out a near-IR spectroscopic follow-up
observation and confirmed two massive red galaxies at the cluster redshift.

Given the the convincing over-density of galaxies and wealth of
imaging and spectroscopic data available in the field, PKS1138 is
an ideal sample to study the environmental dependence of
galaxy properties at this high redshift.
In this paper, we perform an extensive analysis of galaxies around
PKS1138 to quantify the environmental dependence of galaxy evolution
and formation at $z=2.15$.

The layout of the paper is as follows.
We summarize our data in Section 2.
We then  describe details of our method of
fitting spectral energy distributions of galaxies in Section 3.
Before presenting our results, we perform extensive sanity checks in Section 4.
Section 5 presents physical parameters of galaxies obtained from
the fits as a function of environment at $z\sim2$ and
Section 6 discusses implications of our results for galaxy formation.
Finally, we summarize the paper in Section 7.

Unless otherwise stated, we adopt H$_0=70\rm km\ s^{-1}\ Mpc^{-1}$,
$\Omega_{\rm M}=0.3$, and $\Omega_\Lambda =0.7$.
Magnitudes are on the AB system.
We use the following abbreviations: IMF for initial mass function,
SED for spectral energy distribution and SFR for star formation rate.

%-------------------------------------------------------
\section{Data}

We use two data sets for the purpose of this paper.
One is from the (proto-)cluster field PKS1138 to examine galaxies in
an over-density region at $z=2.15$.
The other one is from GOODS for a field counterpart for comparison.
We summarize the two sets of data in this section.

%-------------------------------
\subsection{PKS1138}

The field has been imaged by several instruments ---
$U$-band with LRIS on Keck \citep{zirm08}, $R$ band with FORS on VLT
\citep{kurk00,kurk04a}, $J,\ K_s$ bands with MOIRCS on Subaru \citep{kodama07},
$3.6-8.0\mu m$ with IRAC and $24\mu m$ with MIPS on Spitzer \citep{seymour07}.
In addition to these published data, we used $z$-band data taken with FORS2
on VLT and $H$-band data with SOFI on NTT,
which were obtained as a filler target of other programs
and have not been published elsewhere.
The $z$ and $H$-band data were reduced in a standard manner.
We further supplemented the data set with superb resolution imaging by
the ACS on-board the HST \citep{miley06}.
We retrieved the pipeline reduced $g$ and $I$ band images
from the Hubble Legacy Archive.
Table \ref{tab:data} summarizes the data.

The photometric zero points for the ground-based optical-nearIR images were
obtained from the standard star observations.
As pointed out by \citet{doherty09}, the $J$ and $K_s$ band photometric
zero-points used in \citet{kodama07} were off by $\sim$0.3 mag.
We re-measured the zero points from the standard stars observed
in the same nights and adopted those revised zero points.
For the ACS images, we used the zero points from \citet{sirianni05}.
We have checked all the zero points against stars from \citet{gunn83}.
We convolved the SEDs of stars from \citet{gunn83} with filter responses and
atmosphere and derive synthesized magnitudes of stars.
We then compared the observed sequence of stars with the synthesized sequence
on color-color diagrams with various color combinations.
We found small zero point offsets with respect to the \citet{gunn83} stars and
corrected for them to better match with \citet{gunn83}.
The applied offsets were typically $\lesssim0.05$ mag and the largest
was $0.20$ mag to the $U$-band.

Objects were detected using {\sc sextractor} \citep{bertin96} in the individual
optical and near-IR bands because the seeing sizes vary from band to band and
we did not smooth them to a common seeing.
We used MAG\_AUTO with aperture corrections assuming point sources for the optical
and near-infrared photometry.
For the IRAC images, we performed aperture photometry using the $K_s$-band
image for object detections. We then applied aperture corrections to
obtain the total fluxes.

The aperture corrections were estimated as follows.
We first constructed an average PSF image using bright, unsaturated stars in each band.
Then we randomly distributed PSF objects in the images and repeated
the object detection and photometry.  The differences between the input
magnitudes and measured magnitudes 
were used for the aperture corrections.
We also estimated detection limits for point sources and they are summarized
in Table \ref{tab:data}.  The last column in the table shows the 50\%
detection limits. For images taken with ground-based facilities,
apparent sizes of faint sources are small and magnitude limits for
point sources are a reasonable proxy for extended sources.
For the ACS data, the magnitude limits for point sources are extremely deep
due to the superb resolution.
We measured typical magnitudes of the detected extended sources
where their errors become $0.3$ magnitude
(i.e., $\sim3\sigma$ limits) and adopted them as magnitude limits for
extended sources.

All the photometric errors were estimated from the sky noise in the same aperture
sizes as used for objects.
The photometric catalogs were cross-correlated with the $K_s$-band catalog within
the seeing FWHM of each band, and a $K_s$-band selected catalog was produced.
The final catalog contains $UgRIzJHK_s$ and IRAC $3.5,\ 4.5,\ 5.8\mu m$ photometry and
the Galactic extinction was corrected for in each band using the dust map
from \citet{schlegel98}.
Stars are removed from the catalog based on their compactness and colors
(we use a catalog in which stars are not removed only in Section 4.1).

Note that we have an IRAC $8.0\mu m$ image, but we do not use it
as strong dust emission (e.g., PAH emission at $7.6\mu m$) from
low redshift galaxies fall in this band.
Our models described below do not include dust emission. 
We could in principle run photo-$z$ without the 8$\mu m$ photometry, select
$z\sim2$ galaxies, and then run the full SED fits to the selected galaxies
including the 8um photometry, which does not probe PAH at $z\sim2$.
But, that complicates the error analysis as it is not very straightforward
to quantify how the errors propagate if we use two different sets of templates.

We also have a MIPS image of the PKS1138 field, but we only use it
for a sanity check in Section 4.
We apply a magnitude cut of $K_s<22.5$ ($\sim10\sigma$) to
the catalog to avoid any significant incompleteness effects.
Also, we use only galaxies detected in more than 5 bands to ensure
reliable SED fitting.

%-------------------------------
\subsection{GOODS}

We use data from the GOODS-MUSIC sample \citep{grazian06,santini09}
for a field counterpart of the PKS1138 data.
There is a wealth of data in the GOODS field, but we restrict
ourselves to $U_{35}BVIzJHK_s$ and IRAC $3.6,\ 4.5,\ 5.8\mu m$ data for this work
so that we have almost the same wavelength sampling as the PKS1138 data.
This is a crucial point of this work -- if we used more data
in the GOODS field, then we would not be able to compare PKS1138 and
GOODS due to the different levels of systematic biases in the analyses.
Note that we will use the MIPS photometry just for a sanity check in Section 4.
The data covers a 78 arcmin$^2$ field, which is limited by the $H$-band field coverage.
We apply the magnitude cut of $K_s<22.5$ to the GOODS catalog
and use galaxies with detections in more than 5 bands as done for PKS1138.

To make sure that GOODS is not a peculiar field, we derive
galaxy density in the Subaru XMM-Newton Deep Field (SXDF)
using the photometric data described in Finoguenov et al. (2009),
which covered approximately $\rm 2,000\ arcmin^{2}$.
The density at $z\sim2$ turns out to be similar to the GOODS density;
$0.57\pm0.02\rm\ arcmin^{-2}$ in SXDF and $0.64\pm0.09\rm\ arcmin^{-2}$ in GOODS.
This suggests that GOODS samples a typical Universe at $z\sim2$.

%-------------------------------
\begin{table}
\caption{Photometric data for PKS1138.}
\label{tab:data}
\centering
\begin{tabular}{lccc}
\hline\hline
Band & Instrument & PSF size & 50\% limit\\%  & 20\% limit\\
\hline
$U$  & LRIS       & $1.2''$  & 26.9     \\%   & 27.1\\
$g$  & ACS        & $0.1''$  & 26.9     \\%   & 26.5\\
$R$  & FORS2      & $1.0''$  & 25.3     \\%   & 25.7\\
$I$  & ACS        & $0.1''$  & 26.7     \\%   & 26.6\\
$z$  & FORS2      & $0.9''$  & 24.1     \\%   & 24.7\\
$J$  & MOIRCS     & $0.7''$  & 24.9     \\%   & 25.6\\
$H$  & SOFI       & $0.8''$  & 22.3     \\%   & 23.6\\
$K_s$& MOIRCS     & $0.7''$  & 23.3     \\%   & 23.5\\
$3.6\mu m$ & IRAC & $\sim2.0''$  & 23.1 \\%   & 23.7\\
$4.5\mu m$ & IRAC & $\sim2.0''$  & 22.5 \\%   & 23.6\\
$5.8\mu m$ & IRAC & $\sim2.3''$  & 20.9 \\%   & 21.7\\
$8.0\mu m$ & IRAC & $\sim2.4''$  & 20.6 \\%   & 21.5\\
\hline
\end{tabular}
\end{table}

%-------------------------------------------------------
\section{The SED Fitting Method}

The idea of the work is to use the data from the two fields
with almost the same wavelength coverage and sampling, feed
the data to the same SED fitting code, fit them with the same
set of templates, and discuss {\it relative} differences between the two samples.
We presented our data sets in the last section.
We here describe details of our SED fitting code.
The SED fitting procedure follows the conventional $\chi^2$ minimizing statistics.
We prepare model templates and compare those templates with the observed data.
We then analyze the fits and derive physical parameters out of
the fits such as SFR and dust extinction.
We start with preparing templates and then describe
details of the fitting procedure.

%----------------------------
\subsection{Model Templates}

The model templates are generated using an updated version of
the \citet{bruzual03} population synthesis code,
which takes into account the effects of thermally pulsating AGB stars.
We adopt the Salpeter initial mass function\footnote{
The Chabrier initial mass function
\citep{chabrier03} gives smaller SFRs and stellar masses by a factor of $\sim2$
compared to the Salpeter IMF.
Other than that, it gives the same results as those presented below.
}
\citep{salpeter55} to keep consistency with our previous study \citep{doherty09}
and solar and sub-solar metallicities ($Z=0.02$ and 0.008).
The physical parameters that go into the templates are

\begin{itemize}
\item star formation time scale ($\tau$) assuming
the exponentially declining SFR (i.e., $SFR(t)\propto \exp(-t/\tau)$)
\item optical depth of dust extinction in the $V$-band ($\tau_V$)
\item age, which is the time since the onset of star formation
to the observed epoch.
\end{itemize}

\noindent
We allow $\tau$ to vary between 0 (single burst) and $\infty$ (constant SFR),
and $\tau_V$ between 0 and 10. 
For the single burst model ($\tau=0$), we assume no dust to mimic
the passive evolution.
For the other models, we do not assume any correlation between $\tau$ and $\tau_V$.
We adopt the two component dust extinction model of \citet{charlot00}.
The actual amount of dust applied in the templates is not simply $\tau_V$ but
is dependent on star formation histories.  We measure a $V$-band magnitude of a template
and compare it with the $V$-band magnitude of dust-free template
of the same age, $\tau$, and metallicity.
The difference between the two is $A_V$.
Note that we make fine model grids where $z\sim2$ galaxies populate.
We implement effects of the intergalactic extinction following
\citet{furusawa00}, who used the recipe by \citet{madau95}
and relaxed the allowed range of extinctions, as we explore the $z>2$ Universe.
Also, we use SWIRE AGN templates from \citet{polletta07} as
we might expect an increased fraction of AGNs in the PKS1138 region
\citep{pentericci02,galametz09}.

We convolve the template spectra redshifted to various redshifts
with response functions of the detectors and filters with/without atmosphere
(we do not convolve atmosphere for data from the space)
and generate synthesized fluxes.
We use a logical constraint on redshift that the age of a template spectrum
must be younger than the age of the Universe at that redshift.
We have confirmed that we obtain essentially the same results
if we do not use the constraint as we discuss below.
The library of the synthesized fluxes for all the template is
used to fit the observed SEDs of galaxies as detailed below.
We use 2 million templates in total for both PKS1138 and GOODS.

%----------------------------
\subsection{Fitting procedure}

We use the conventional $\chi^2$ minimizing statistics to
fit the models to the observed data.
The fit itself is performed in fluxes, not in magnitudes.
Non-detections are treated as zero flux with flux errors from
the magnitude limits (here we take the 50\% detection limits).
In this way, we avoid using upper limits in the fits and
keep the $\chi^2$ statistics simple.
We impose a prior constraint that all galaxies have stellar masses
lower than $6\times10^{11} \rm M_\odot$.
This is a reasonable constraint -- 
the most massive galaxies at $z\sim2$ have
$\sim5\times10^{11}\rm M_\odot$ \citep{kriek08}.

The best fitting template gives photometric redshift and
physical parameters of galaxies. 
To be specific, we measure $z_{phot}$, $\tau$, $\rm A_V$, age, SFR,
and stellar mass from the fits.
If the best fitting template is an AGN template, then we can not extract
these parameters except for $z_{phot}$ because there is an non-negligible
contribution of AGN to the overall SED.
We only use their photometric redshifts.

\begin{figure}
\centering
\includegraphics[width=8cm]{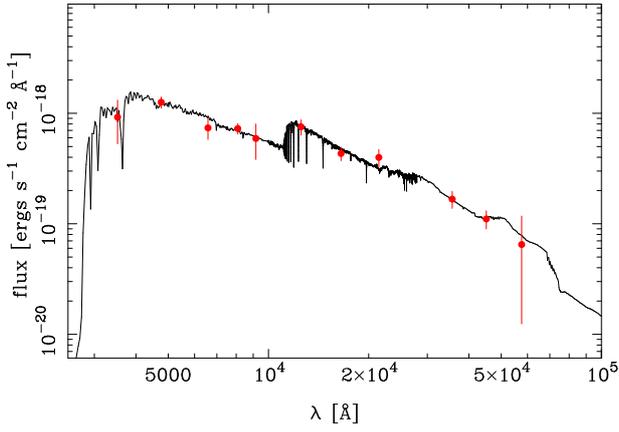}
\caption{
Sample SED fit.  The points are the observed photometry
and the spectrum is the best-fitting template spectrum.
For this object, we obtain $z_{phot}=2.00^{+0.16}_{-0.49}$,
age$=0.32^{+0.28}_{-0.24}$ Gyr, $\tau=1.0^{+\inf}_{-0.9}$,
$A_V=0.81^{+0.51}_{-0.27}$ mag., SFR$=78^{+31}_{-63}\rm\ M_\odot\ yr^{-1}$,
and stellar mass of $2.4^{0.5}_{-1.2}\times 10^{10}\rm M_\odot$
from the fit.
}
\label{fig:sed_eg}
\end{figure}

An uncertainty on each parameter is estimated by marginalizing over
the parameter of interest and taking an interval of $\Delta \chi^2=1$.
However, all the parameters are correlated and this is just a rough error estimate.
It is not easy to handle all the co-variances correctly, but fortunately,
accurate random error estimates are not crucial for our purpose.
It is because we discuss relative, systematic differences between the two data sets.
We should have the same level of co-variances between the parameters
as we use the same code, templates and have almost the same wavelength sampling of SEDs.
We therefore take this simple error estimate, but we should bear in mind
that our errors are not accurate.
Fig. \ref{fig:sed_eg} shows an example of our SED fits.

%-------------------------------

%-------------------------------------------------------
\section{Sanity Checks}

Before we present our results, we perform a few sanity checks
to make sure that our results are robust.
Here we discuss (1) photometric zero points,
(2) accuracy of our photometric redshifts,
(3) accuracy of the SFR estimates from the SED fits,
and (4) secondary burst models.

%----------------------------
\subsection{Photometric Zero Points}

We cannot avoid random errors in the photometric zero points,
but it is important to make sure that we do not have any
systematic zero point offsets as a function of wavelength, which can cause
color stretches in the color space.
For example, if the zero points are systematically fainter in the blue bands,
we might get higher extinction and/or older ages from the SED fits.

\begin{figure}
\centering
\includegraphics[width=8cm]{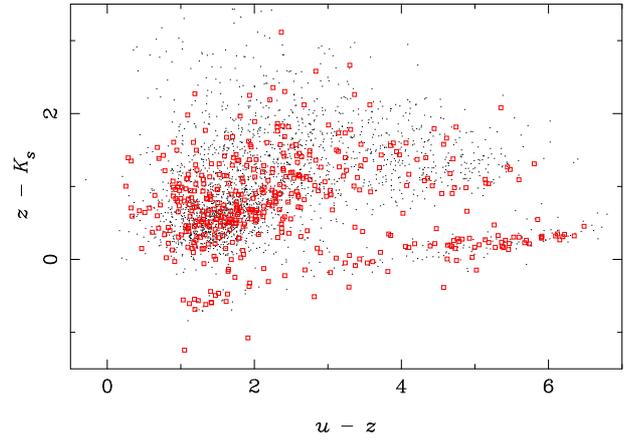}
\caption{
$z-K_s$ plotted against $u-z$ for objects brighter than $K_s=22.5$.
The open squares and dots show objects in PKS1138 and GOODS, respectively.
Note the sequence of stars in the lower part of the plot.
}
\label{fig:col_col}
\end{figure}

The easiest way to check the color stretch is to look at the sequence
of stars in color-color diagrams.
Fig. \ref{fig:col_col} is one such plot.
The plot shows $z-K_s$ vs. $u-z$ of the detected objects brighter than
$K_s=22.5$ in the two samples.
One can easily identify the stellar sequence in the lower part of the plot.
There is no strong systematic offset between the sequence of
stars in the PKS1138 field and that in the GOODS field.
We have also checked various color combinations and found that the
offsets are $<0.1$ mags.
To accommodate with any possible zero point errors,
we add 0.1 mag in the quadrature to all the magnitude uncertainties
in both PKS1138 and GOODS catalogs before performing the SED fits
described in the last section\footnote{
The addition of random errors is not an ideal way to remedy
the systematic errors.  But, we apply the additional errors
to make sure that the systematic errors do not dominate
the overall error budget.
The price we have to pay is that the additional errors
slightly degrade the accuracy of photometric redshifts.
}.

%-------------------------------

%-------------------------------
\begin{figure}[h]
\centering
\includegraphics[width=8cm]{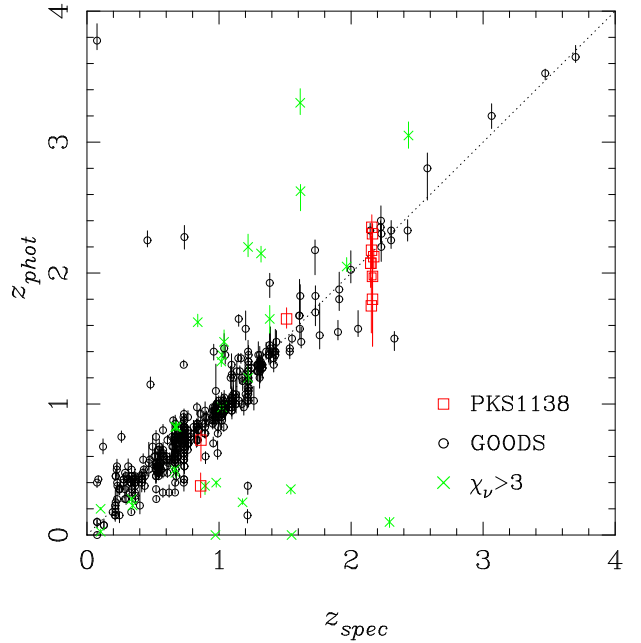}
\caption{
$z_{phot}$ plotted against $z_{spec}$.
The dotted line shows the $z_{spec}=z_{phot}$ relation.
The squares and circles are $K_s<22.5$ galaxies in PKS1138 and GOODS, respectively.
The crosses show objects with $\chi^2_\nu>3$ both for PKS1138 and GOODS.
}
\label{fig:photoz_specz}
\end{figure}

%----------------------------
\subsection{Photometric Redshifts}

Another good check will be to look at the accuracy of photometric redshifts.
We want to make sure that we can cull the majority of the galaxies
at $z\sim2$ with a reasonably small amount of fore-/background contamination.

Fig. \ref{fig:photoz_specz} compares our photometric redshifts ($z_{phot}$)
with spectroscopic redshifts ($z_{spec}$).
The spectroscopic redshifts in PKS1138 are from \citet{pentericci00},
\citet{kurk04a}, \citet{croft05}, and \citet{doherty09} and those in GOODS are from
the GOODS-MUSIC catalog (\citealt{grazian06} and references therein).
The correlation is reasonably good, although there are a number of
galaxies with wrong photo-$z$.
We find that galaxies with poor fits (the reduced $\chi^2$ larger than 3)
often have wrong photometric redshifts.
We remove those galaxies from the main analysis.
Our results remain unchanged if we do not apply this $\chi^2_\nu$ cut 
as we quantify later.

A fraction of photo-$z$ outliers ($|z_{spec}-z_{phot}|/(1+z_{spec})>0.2$)
is 5\% in GOODS over the entire redshift range.
Our spectroscopic sample for PKS1138 is strongly
skewed at particular redshifts, and we do not quote the numbers
for comparison here.
The photo-z selection criterion for the main analysis is a trade off
between the completeness and contamination.
The selection adopted for the main analysis, $z_{phot}=2.15$ within $2\sigma$,
is primarily motivated by the rate of recovering the spectroscopically
confirmed PKS1138 members
(it recovers 6 out of 7 confirmed PKS1138 members brighter than $K_s=22.5$).
We have confirmed that our results presented below do not change
if we change the selection criterion within a reasonable range
(e.g., $|z_{phot}-2.15|<0.2$).
The contamination of fore-/background galaxies in our sample used
in the main analysis is 36\% in GOODS,
which is estimated from the spectroscopically confirmed galaxies at
$|z_{spec}-2.15|>0.3$ with photometric redshift consistent with $z_{phot}=2.15$
within $2\sigma$. 
Note that the numbers quoted here should be considered rough estimates
because the spectroscopic redshifts are collected from various surveys
and the sample is quite heterogeneous.
As we will show later, contamination due to wrong photometric redshifts
(e.g., foreground galaxies at $z_{spec}\sim1$ but with $z_{phot}\sim2$)
do not strongly affect our conclusions.
We note in passing that 6 out of 50 galaxies in PKS1138 and 8 out of
50 galaxies in GOODS that are used in the main analysis are
spectroscopic members.

It is also important to verify that the parameters derived
from the SED fits are not strongly affected by the uncertainties in
the photometric redshifts.
We take into account the random errors when we discuss
differences between PKS1138 and GOODS in the main analysis,
but one may worry that photometric redshifts may introduce systematics
in the SED fits and in the derived parameters.

\begin{figure*}[t]
\centering
\includegraphics[height=7cm]{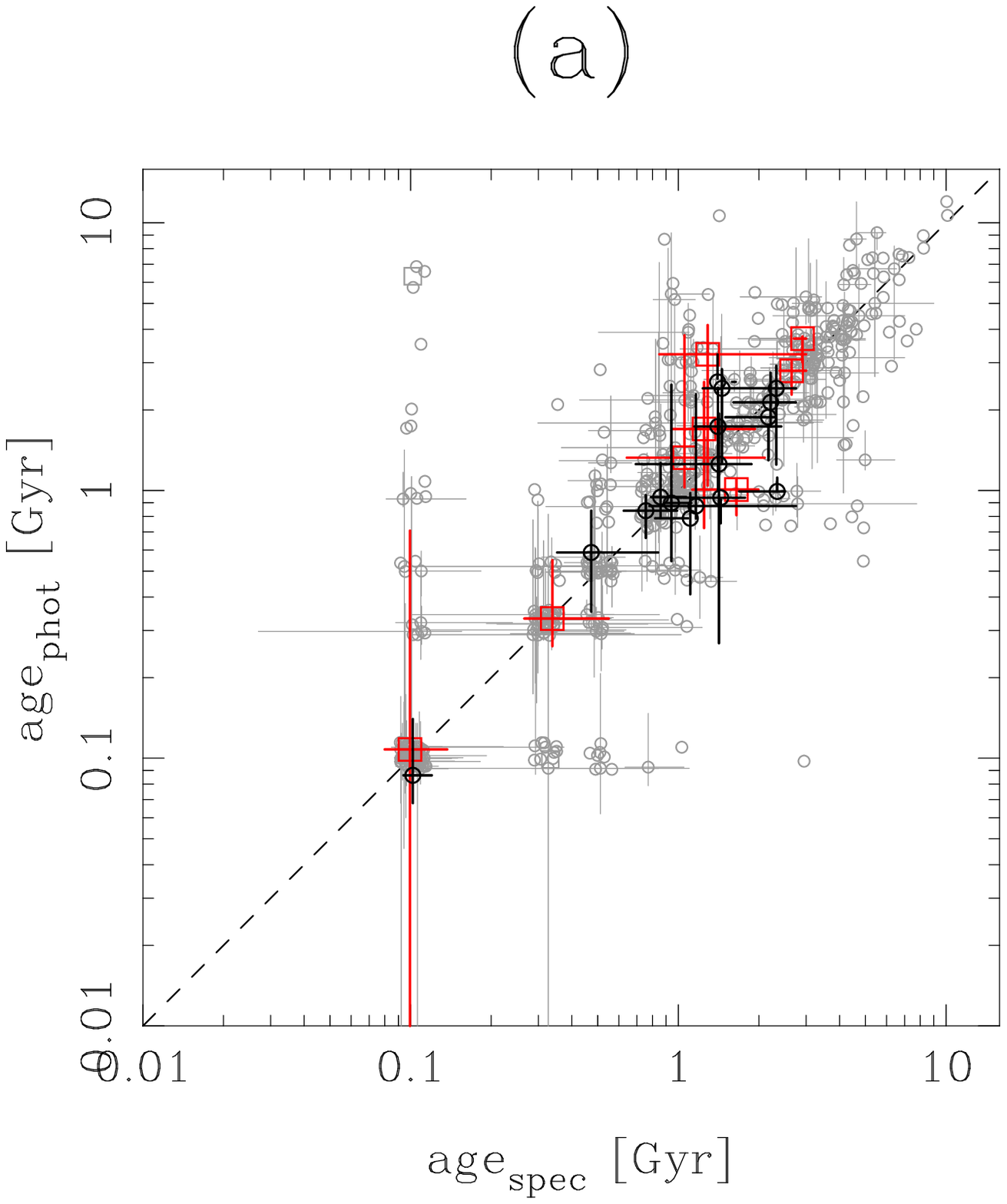}\hspace{0.5cm}
\includegraphics[height=7cm]{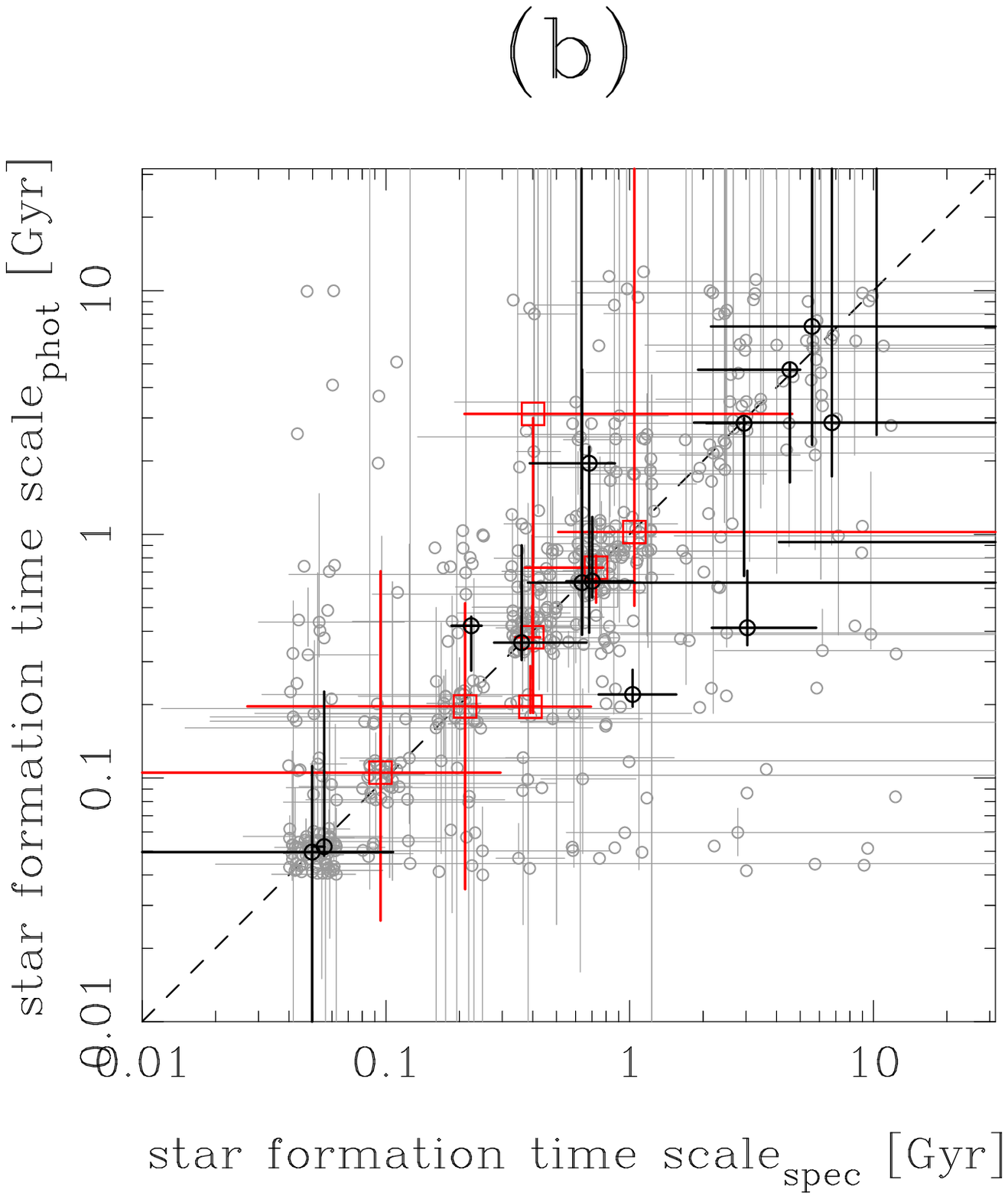}\\\vspace{0.5cm}
\includegraphics[height=7cm]{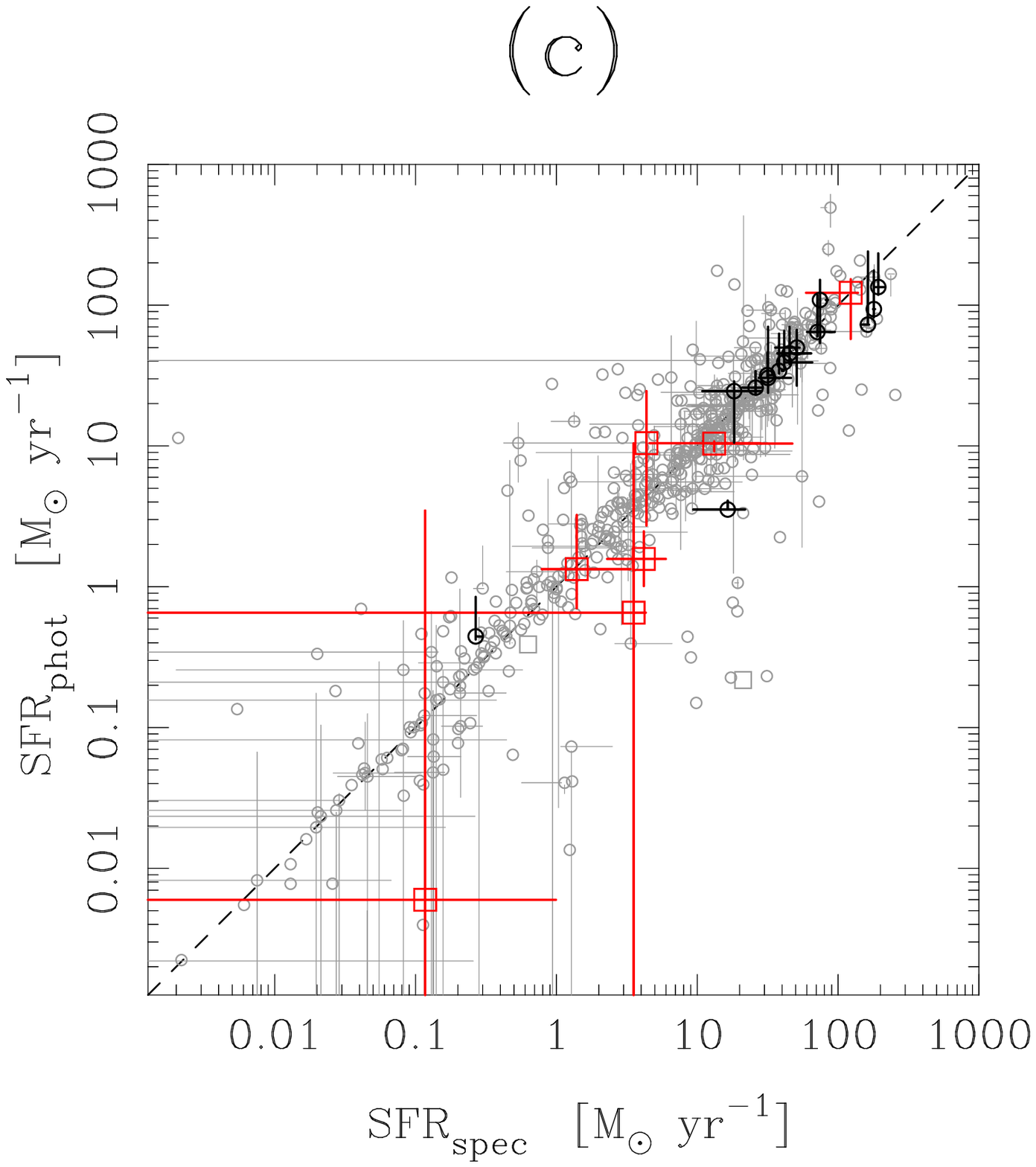}\hspace{0.5cm}
\includegraphics[height=7cm]{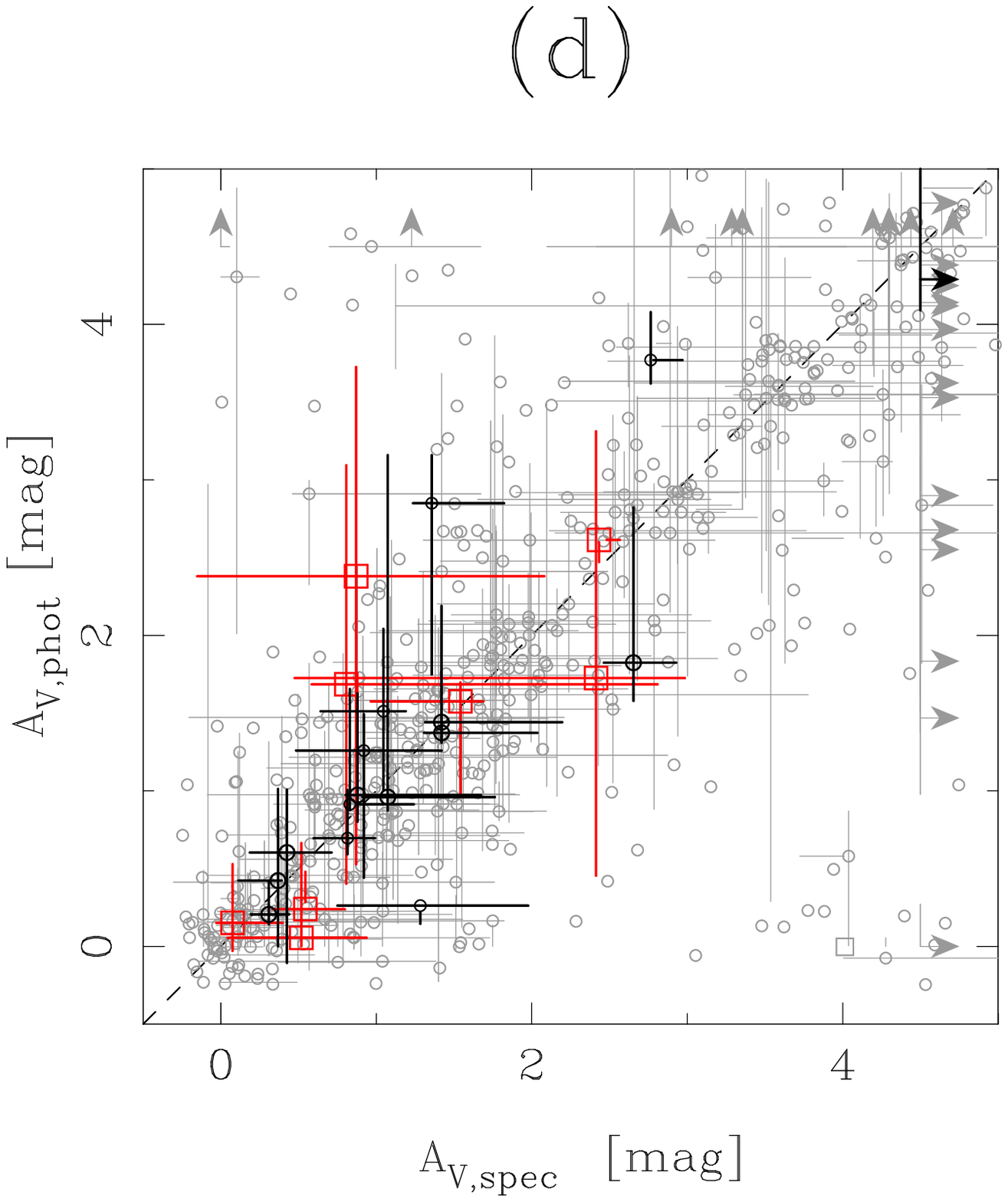}
\caption{
Comparisons between the parameters derived at $z_{spec}$ and $z_{phot}$.
The parameters plotted in the x-axis are derived by fixing
redshifts at $z_{spec}$, and those in the y-axis are derived
using redshift as a free parameter.
The panels show (a) age, (b) star formation time scale $\tau$,
(c) SFR, and (4) dust extinction in $\rm A_V$.
The squares and circles are objects in PKS1138 and GOODS, respectively.
The thick symbols show the spectroscopic members
(i.e., galaxies at $z\sim2.15$) and pale symbols show non-members.
For clarity, we plot the error bars for every 5 objects
for non-members.
The points are randomly shifted by a small amount to avoid overlapping.
}
\label{fig:comp_param}
\end{figure*}

To quantify this, 
we fit SEDs at redshifts fixed at $z_{spec}$ for the spectroscopic objects.
Fig. \ref{fig:comp_param} compares the parameters derived at $z_{spec}$
and those with redshifts as a free parameter.
We observe a good correlation between them.
In particular, there is no strong systematic offsets
as shown in the top panel of Table \ref{tab:med_diff}, where we show
the median differences between the physical parameters derived at $z_{phot}$
and those at $z_{spec}$  for the spectroscopic members at $z_{spec}\sim2.15$.
The values in the brackets show the rate at which the two parameters agree within $1\sigma$.
Spectroscopic redshifts have narrow errors on all the parameters, but
photometric redshifts do not introduce any systematics.
We will focus on systematic differences between the two samples
in the next section and all the differences we observe are larger
than the offsets listed in Table  \ref{tab:med_diff}.
Therefore, the uncertainties in the photometric redshifts
do not affect our results.

We will not use parameters derived at $z_{spec}$ in the main analysis
because we would like to show what parameters the contaminant galaxies
(i.e., objects with wrong photometric redshifts) typically have
and to show that they do not affect our conclusions.
Also, the spectroscopic samples are strongly biased towards particular
types of galaxies and we do not want
to put a weight on them.  For example, many of the spectroscopically
observed galaxies in PKS1138 are either emission line objects or
X-ray sources \citep{pentericci00,kurk04a} and they are not necessarily
typical galaxies.
We use photometric redshift as a free parameter for all the objects
so as not to introduce any biases.
We have confirmed that our results do not change
if we use the parameters derived at $z_{spec}$ where available,
as expected from Fig. \ref{fig:comp_param}.

Finally, we show that the logical age constraint - the age of
a model template must be younger than the universe at a given redshift -
does not change our results.
The middle panel of Table  \ref{tab:med_diff} quantifies the effect
of the age constraint on the derived parameters.
We observe no systematic differences and the derived parameters of
most galaxies agree within $1\sigma$.
We can therefore safely apply the constraint in the main analysis.
The reason why the observed offset is zero is because
the model grids are discrete and many galaxies remain in the same grid.
SFRs do not have grids as they are normalized by the observed fluxes
of galaxies, but most galaxies have the same best-fitting
templates and the median SFRs remain the same.

%-------------------------------

%-------------------------------
\begin{table}
\caption{
%Median differences between the physical parameters between
%{\it top:} those derived at $z_{phot}$ and those at $z_{spec}$,
%{\it middle:} those with/without the logical age constraint, and
%{\it bottom:} those with the secondary burst models and those with the $\tau$ models.
Differences in the physical parameters derived with other models
or constraints than the fiducial ones.
}
\label{tab:med_diff}
\centering
\begin{tabular}{lrr}
\hline\hline
Parameter                    & PKS1138         & GOODS\\
\hline
age [Gyr]                    & $+0.13$ (75\%)  & 0    (81\%)\\
SF time scale [Gyr]          & 0       (100\%) & 0    (70\%)\\
SFR [$\rm M_\odot\ yr^{-1}$] & $-0.08$ (88\%)  & $-0.6$ (81\%)\\
$\rm A_V$ [mag]              & 0       (100\%) & 0     (88\%)\\
%----no age constraint
\hline
age [Gyr]                    & 0       (92\%)  & 0    (96\%)\\
SF time scale [Gyr]          & 0       (90\%)  & 0    (100\%)\\
SFR [$\rm M_\odot\ yr^{-1}$] & 0       (76\%)  & 0    (92\%)\\
$\rm A_V$ [mag]              & 0       (80\%)  & 0    (97\%)\\
%---secondary burst models
\hline
age [Gyr]                    & 0      (70\%)   & 0    (79\%)\\
SF time scale [Gyr]          & 0      (72\%)   & 0    (80\%)\\
SFR [$\rm M_\odot\ yr^{-1}$] & $-0.05$ (56\%)  & 0    (73\%)\\
$\rm A_V$ [mag]              & $-0.50$ (58\%)  & $-0.75$ (37\%)\\
\hline
\end{tabular}
\tablefoot{
The top panel shows the median differences between the parameters derived
at $z_{phot}$ and those at $z_{spec}$.  The numbers in the parenthesis
shows the fraction at which the two parameters agree within $1\sigma$.
The middle panel shows the differences we obtain if we do not
use the logical age constraint.
The numbers in the bottom panel mean the median differences between
parameters from the secondary burst models and those from the fiducial
$\tau$ models.
}
\end{table}

%-------------------------------
\begin{figure}
\centering
\includegraphics[width=8cm]{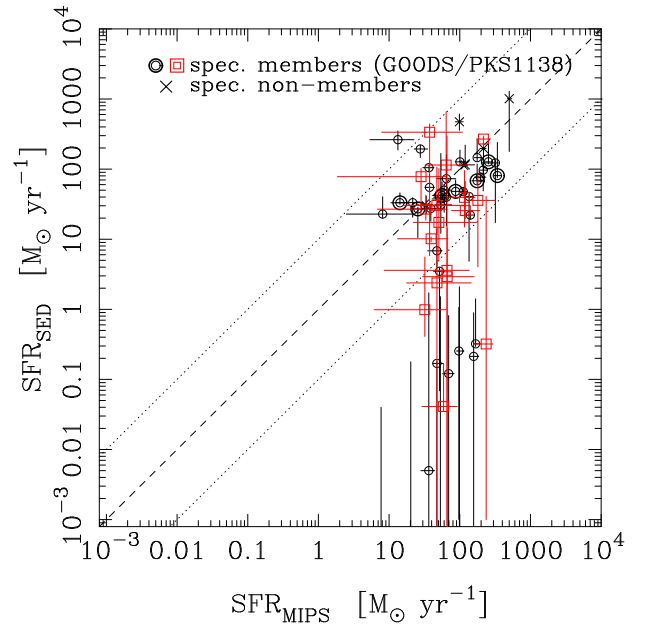}
\caption{
SFR from the SED fits plotted against SFR from the MIPS data in GOODS (circles),
MIPS in PKS1138 (squares)
for galaxies around the radio galaxy redshift ($z=2.15$).
The double circles and squares show the spectroscopically confirmed members
in GOODS and PKS1138, respectively, and the crosses are confirmed
outliers.
Here we plot only galaxies used in the main analysis.
The dashed line shows $\rm SFR_{SED}=SFR_{MIPS}$ and the dotted lines
show 1 dex offsets.
}
\label{fig:sfr_comp}
\end{figure}

%----------------------------
\subsection{SFR from SED Fits}

Next, we check the accuracy of the SFRs from the SED fits.
There are many ways to derive SFRs of galaxies, but here we take
those from the warm dust emission probed by the MIPS observations.
We take the conversion from the MIPS flux to the total IR flux
by \citet{reddy06} and derive SFRs using the formula given by \citet{kennicutt98}.
We have a deep MIPS $24\mu m$ image in both PKS1138 \citep{seymour07} and GOODS
\citep{santini09}, reaching down to $\sim 40\rm M_\odot\ yr^{-1}$
and  $\sim 15\rm M_\odot\ yr^{-1}$, respectively, at $z=2.15$.
It should be possible to measure SFRs from various spectroscopic
observations carried out in the GOODS field, but the spectra
are not always public.

We plot in Fig. \ref{fig:sfr_comp} SFRs from the SED fits against
SFRs from MIPS.
%/H$\alpha$/Ly$\alpha$.
%\sout{
%We supplement the plot with Ly$\alpha$ and H$\alpha$ emitters
%in the PKS1138 confirmed by \citet{pentericci00}, \citet{kurk04a},
%and \citet{doherty09}.
%Note that the SFRs from the Ly$\alpha$ and H$\alpha$ are not
%corrected for dust extinction and therefore they are lower limits
%of the true SFRs (and they are plotted as such).
%SFRs from Ly$\alpha$ and H$\alpha$ lines are also obtained using the formula
%from \citet{kennicutt98}.
%}
Note that we plot only galaxies that are detected in MIPS and are
at $z_{phot}=2.15$ within $2\sigma$ used
in the main analyses presented in the following sections.
For the GOODS sample, the agreement between $\rm SFR_{SED}$ and
$\rm SFR_{MIPS}$ is relatively good, although there are some outliers.
Roughly $70\%$ of the galaxies have consistent SFRs from the SED fits and
MIPS photometry within 1 dex at $1\sigma$.
Note that $\rm SFR_{MIPS}$ itself must have a significant error given
the large scatter in mid-far IR SEDs of galaxies \citep{elbaz02}.

The PKS1138 sample has a similar rate of catastrophic failures to the GOODS sample --
6 out of 24 galaxies ($\sim25 \%$) plotted fall below MIPS SFRs by more than 1 dex,
although only one of them is inconsistent at $1\sigma$.
SFRs for the PKS1138 galaxies have larger errors compared to the GOODS SFRs.
This is because some of our optical images in the PKS1138 are shallow
compared to the GOODS images.
The optical images probe rest-frame UV at $z\sim2$ and they are
crucial to pin down accurate SFRs.

We have visually inspected the galaxies with $SFR_{SED}+\sigma(SFR)<10\rm M_\odot\ yr^{-1}$
(i.e., the upper error is lower than $10\rm M_\odot\ yr^{-1}$)
used in the main analysis on the MIPS images.
75\% (9 out of 12) of galaxies with low SFRs in GOODS
are detected in MIPS and have $SFR_{MIPS}\gtrsim40\ \rm M_\odot\ yr^{-1}$.
The other three are also detected but have $<20\ \rm M_\odot\ yr^{-1}$.
This high fraction of the MIPS detection illustrates the fundamental
difficulty in distinguishing dusty starbursts from passive galaxies.
But, this can be due to wrong photometric redshifts, AGN contamination,
or object blending/confusion in the MIPS image due to the large PSF.
We have looked at X-ray point sources in the field \citep{lehmer05}
and found that none of the low SFR galaxies are detected.
We note in passing that 3 out of 50 GOODS galaxies used in the main analysis
are detected in X-rays, but they do not affect our results.

In contrast to GOODS, the PKS1138 galaxies used in the main analysis
show a difference between low and high SFR galaxies -- 
only $\sim20\%$ of galaxies (4 galaxies out of 21) with
$SFR_{SED}+\sigma(SFR)<10\rm M_\odot$ are detected in the MIPS image.
Our MIPS image reaches $SFR_{MIPS}\sim40\rm M_\odot$ at $z=2.15$
and if all the low SFR galaxies in PKS1138 are forming stars
at similar rates to the GOODS galaxies, they should have been detected.
We will show that PKS1138 galaxies have lower SFRs than GOODS galaxies
in the next section and the observed lower frequency of the MIPS detections
in PKS1138 gives a strong support to this result.
We note that \citet{galametz09} suggested an increased fraction of AGNs
in proto-clusters and some of the PKS1138 galaxies with
low SFRs may host an AGN, boosting the mid-IR flux.
In fact, one of the low SFR galaxies that is used in the main analysis
is detected in X-rays \citep{pentericci02}.
This is the only object detected in X-rays and used in the main analysis
and this object does not affect our result in any way.

We admit that the correlation between $SFR_{SED}$ and those measured
from other methods is not particularly good, and we do not attempt to
discuss precise SFRs of individual galaxies.
Instead, we classify galaxies into two coarse classes by their SFRs.
We define high SFRs as $SFR_{SED}>10\rm M_\odot\ yr^{-1}$,
and low SFRs are lower than that.
We do not try to further subdivide the SFRs.
However, we will later introduce another class - starbursts - as
$SFR_{SED}>100\rm M_\odot\ yr^{-1}$ just to illustrate starbursting
populations at the time of cluster formation at very high redshifts.

%----------------------------
\subsection{Secondary burst models}

As we discuss later, cluster galaxies may have experienced many mergers
in the past and the $\tau$ models may not reproduce their star formation
histories very well.
We have implemented a secondary burst on top of the $\tau$ models
and checked how the results change.
We assume an instantaneous burst and add a secondary burst
either at 0.2, 0.5, 1, 2, and 5 Gyr after the onset of star formation
and the burst adds either 10\% or 50\% of stellar mass of the galaxy.

The bottom line of this exercise is that our conclusions remain unchanged.
%The only small difference is that GOODS galaxies have slightly younger ages.
%Most GOODS galaxies at $z=2.15$ are still actively forming stars and
%it is difficult to constrain their recent star formation histories because
%their light is dominated by newly born young stars.
%It seems that the burst models can reproduce these galaxies with
%younger ages than the tau models, although we are not sure
%if this is physically correct.
The bottom panel of Table \ref{tab:med_diff} quantifies the difference
between the secondary burst models and the $\tau$ models.
We observe no strong differences in age and $\tau$.
% but we should note that
%interpretations of age and $\tau$ are not very straightforward due
%to the secondary burst.
The agreement of SFRs between the two models is not particularly good,
but our results remain unchanged because the systematic differences are very small.
The dust extinction shows a poor agreement and a systematic offset in $\rm A_V$.
But, the amount of the offset is similar in PKS1138 and GOODS, and our
conclusion will therefore remain the same as we discuss only relative
differences between the two samples.

%On the other hand, we did not observe any significant changes in the properties
%of the PKS1138 galaxies.
%Most PKS1138 galaxies are not actively forming
%stars and it is easier to constrain recent star formation histories
%of such galaxies.  As shown later, massive galaxies in PKS1138
%have red colors and low SFRs.  If they experienced a burst in a recent
%past, we would not have observed such galaxies because it takes
%time ($\sim1$ Gyr) for them to settle down on the red sequence.
%Mergers that happened long ago could effectively be absorbed in
%star formation time scale (see discussion in Section 6),
%and this might be the reason why the age distribution of the PKS1138 galaxies
%does not change.

%In any case, we find that the secondary bursts do not strongly
%change our primary conclusions.
The secondary burst models are probably more realistic than
the simple $\tau$ models.
However, the addition of the secondary burst raises the number of free
parameters in our SED fitting from 5 to 7 (or 8 if we allow the star formation
time scale of the secondary burst to vary), while we only have 11 broad band
photometry points.  As this introduces too many degeneracies between
the parameters and the secondary burst complicates the interpretation
of the results, we will consider only the $\tau$ models for our main analysis in
the following sections.

%-------------------------------------------------------
\section{Results}

We now move on to present the results from the SED fits.
We stress again that we use the two samples with almost the same
wavelength samplings, feed them to the same code with
the same templates, and discuss the relative differences
between the two samples.

In what follows, we study galaxies with $K_s<22.5$,
detected in more than 5 bands, and being consistent with $z_{phot}=2.15$
within $2\sigma$ (note the PKS1138 radio galaxy is at $z=2.15$).
We exclude galaxies with poor fits  ($\chi^2_\nu>3$)
as they likely have wrong photometric redshifts as discussed in the last section.
We take the same sample definition in both PKS1138 and GOODS,
so that we can make a fair comparison.
The physical parameters we focus on in this section are age, star formation
time scale ($\tau$), dust extinction, and SFR.
We first present evidence that the PKS1138 is indeed an over-density region
as suggested by the earlier studies.
We then move on to discuss physical properties of galaxies from the SED fits
as a function of environment.

%-------------------------------
\subsection{Over-density of galaxies in PKS1138}

\begin{figure*}[tbh]
\centering
\includegraphics[height=10cm]{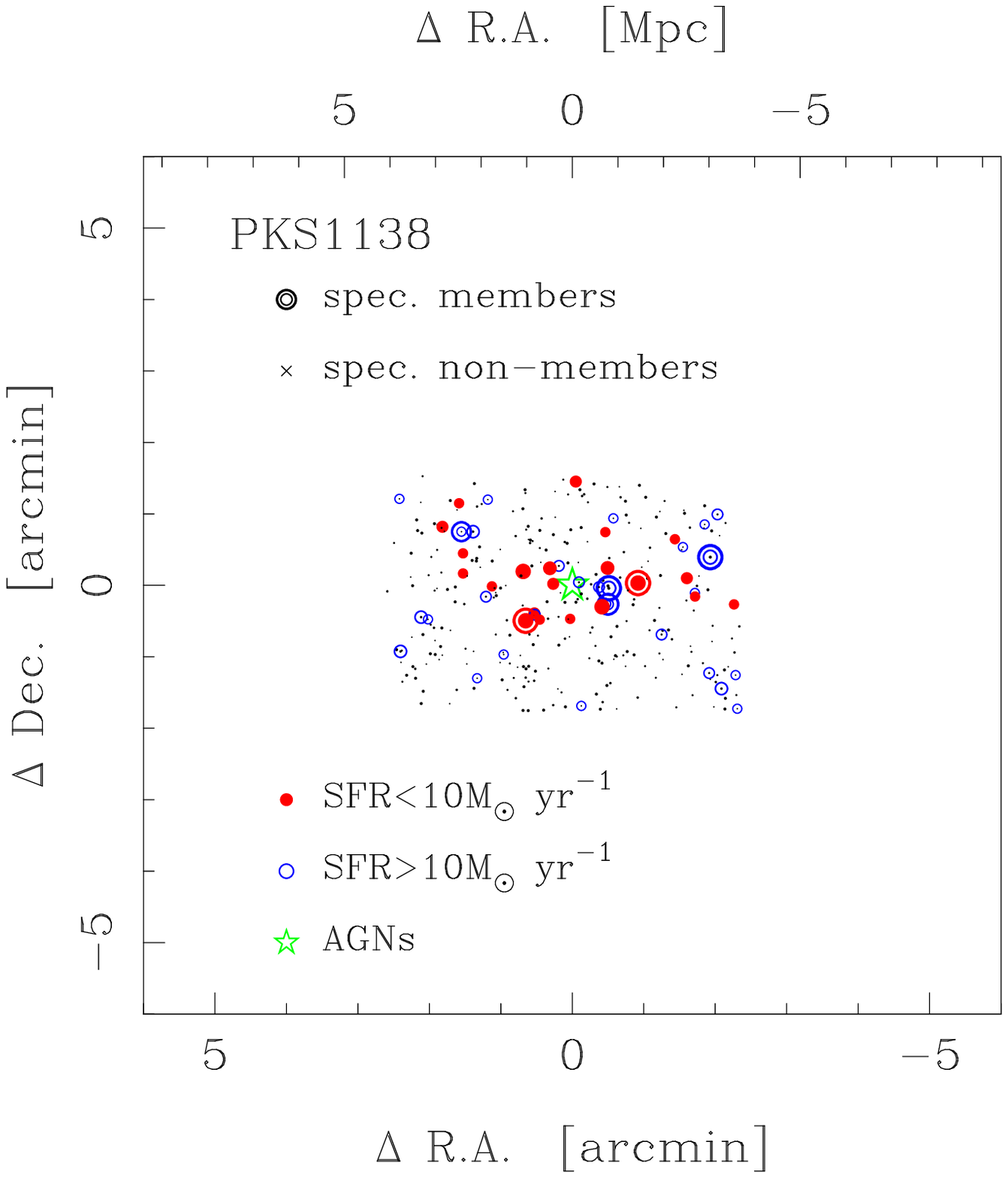}
\includegraphics[height=10cm]{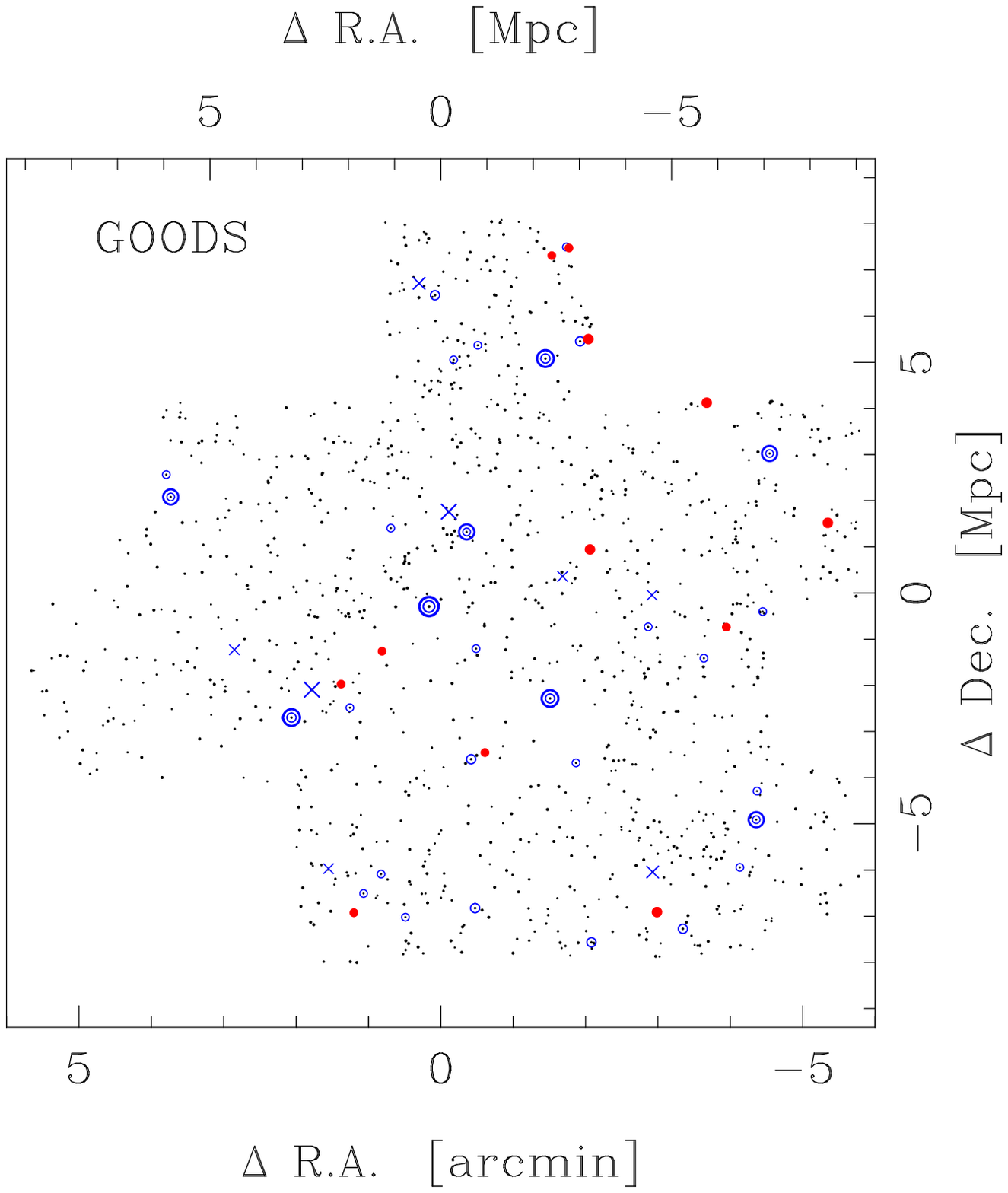}
\caption{
Distributions of galaxies with $K_s<22.5$ at $z_{phot}=2.15$ within $2\sigma$
in the PKS1138 field (left) and GOODS field (right).
The filled circles show galaxies with low SFRs.
To be conservative about red galaxies, we define low SFR galaxies as
$\rm SFR+\sigma(SFR)<10\ M_\odot\ yr^{-1}$ (i.e., the upper error is
lower than 10 $\rm M_\odot\ yr^{-1}$).
The open circles are galaxies with high SFR.
The stars show galaxies fit with AGN templates.
The double circles and crosses mean spectroscopically confirmed
members and outliers, respectively.
The sizes of the symbols correlate with the $K_s$ band luminosity.
The top and right axes show comoving scales at $z=2.15$.
}
\label{fig:lss}
\end{figure*}

Fig. \ref{fig:lss} compares the distributions of galaxies around the radio
galaxy redshift in the PKS1138 and GOODS fields.
We could not obtain a good enough photometric redshift for
the PKS1138 radio galaxy, $z_{phot}=2.38^{+0.08}_{-0.10}$
(i.e., it is 2.3 $\sigma$ away from the spectroscopic redshift)
and it goes out of our main sample.
But, the radio galaxy is a powerful AGN and our photometric redshift
is relatively good for such an extreme object.
We show the radio galaxy in Fig.  \ref{fig:lss} just to illustrate the location
of the radio galaxy.

%-------------------------------

The galaxies around PKS1138 are arrayed in a filamentary structure
in the east-west direction, and the radio galaxy is located at the center.
There is an over-density of galaxies around the central radio galaxy.
Interestingly, many of the galaxies around the radio galaxy show relatively
low SFRs.
On the other hand, the galaxy distribution in the GOODS field is quite sparse
and there is no concentration of low SFR galaxies.
The surface densities of galaxies ($K_s<22.5$) are $3.12\pm0.44\rm\ arcmin^{-2}$
and $0.64\pm0.09\rm\ arcmin^{-2}$ in PKS1138 and GOODS, respectively,
suggesting an over-density of a factor of $\sim 5$ in PKS1138,
which is in rough agreement with \citet{kurk04a} who found a factor of
$\sim4$ over-density of H$\alpha$ galaxies.
The over-density is more pronounced for bright galaxies ;
$0.75\pm0.22\rm\ arcmin^{-2}$ and $0.08\pm0.03\rm\  arcmin^{-2}$ for $K_s<21.5$
(i.e., a factor of 9).
All the above numbers for PKS1138 are averaged over the entire probed field
and the over-density can be further pronounced if we take only
the central part of the PKS1138 field --
the density of PKS1138 galaxies within 1\arcmin\ from the radio galaxy is
$10.2\pm1.8\rm\ arcmin^{-2}$ for $K_s<22.5$ (i.e., a factor of 16)
and $1.59\pm0.71\rm\ arcmin^{-2}$ for $K_s<21.5$ (i.e., a factor of 20).

One may wonder that nearby fore-/background structure may be contaminating
the structure we observe in PKS1138.  But, the contamination should be small
because the observed galaxy distribution is similar to the distribution
of H$\alpha$ emitters at the radio galaxy redshift \citep{kurk04a} --
weak concentrations of galaxies on the West and South-East of
the radio galaxy are both seen in the distributions of photo-$z$ selected
and H$\alpha$ selected galaxies.
Also, structures at different redshifts are unlikely to %conspire to
make a coherent structure around the radio galaxies with
a concentration of low SFR galaxies.
\citet{zirm08} found a concentration of red galaxies selected
in $J-H$ colors around the very vicinity of the radio galaxy ($\lesssim10$\arcsec),
we do not observe such a strong concentration of red galaxies,
but it might be a concentration of faint galaxies (we use $K_s<22.5$,
while they used $H<24.5$, roughly corresponding to $K_s\lesssim24$).

\begin{figure}
\centering
\includegraphics[width=8cm]{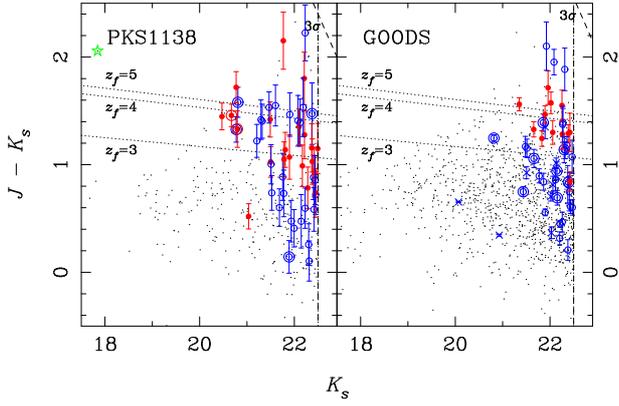}
\caption{
$J-K_s$ plotted against $K_s$.
The dots are all the detected galaxies in each field and large symbols
are galaxies around the cluster redshift ($z_{phot}=2.15$ within $2\sigma$).
The meanings of the symbols are the same as in Fig. \ref{fig:lss}.
Namely, the filled circles and open circles are galaxies
with low and high SFRs, and the stars are AGNs.
The red bright galaxy with $K_s=18$ in the left panel is the PKS1138 radio galaxy.
The vertical dashed lines show $K_s=22.5$.
and the slanted lines show the $3\sigma$ limiting colors.
The dotted lines show the location of the model red sequence
formed at $z=3$, 4 and 5 from bottom to top.
}
\label{fig:cmd}
\end{figure}

Fig. \ref{fig:cmd} plots color magnitude diagrams of galaxies in the two fields.
We model the location of the red sequence using the updated \citet{bruzual03}
population synthesis models following the procedure described in \citet{lidman08}.
The most striking trend is that there is a hint of the red sequence,
which is a ubiquitous feature of galaxy clusters at lower redshifts,
at $K_s\lesssim21$ in PKS1138.
The bright red galaxies are around the model red sequence formed at $z_f\sim4$.
This suggests that these red galaxies formed the bulk of their stars around
that redshift.
The possible red sequence in PKS1138 is populated half by galaxies
with low SFRs ($<10\rm\ M_\odot\ yr^{-1}$)
and half by galaxies with high SFRs, suggesting that
the red sequence is being formed. 
We note that spectroscopically confirmed contaminant galaxies 
tend to be blue, and thus the possible red sequence
in PKS1138 is not due to contamination from fore-/background galaxies.
In contrast to PKS1138, there is no clear sign of a red sequence in the GOODS field,
and there are very few galaxies brighter than $K_s=21$.

\begin{figure}
\centering
\includegraphics[width=8cm]{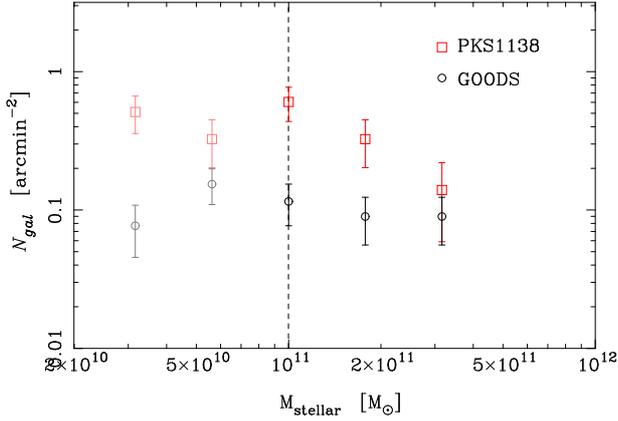}
\caption{
Stellar mass functions of galaxies in PKS1138 (open square) and GOODS (open circle).
The vertical dashed line shows our stellar mass limit of $10^{11}\rm M_\odot$.
The error bars show the Poisson errors.
}
\label{fig:smf}
\end{figure}

To further illustrate the over-density of PKS1138, we plot stellar
mass functions in Fig. \ref{fig:smf}.
We do not derive volume densities of galaxies as we are probably
looking at a coherent structure in redshift in PKS1138, whose size
we do not yet know precisely.
We simply derive a surface density.
The over-density of PKS1138 is evident -- PKS1138 hosts galaxies
more densely than GOODS.
We will focus on galaxies more massive than $10^{11}\rm M_\odot$
in the next section, for which our samples are nearly complete
in both PKS1138 and GOODS.

To sum up, we observe the significant over-density of galaxies and
the forming red sequence in the PKS1138.
This suggests that there is a proto-cluster (or perhaps a real cluster)
around the radio galaxy.
But, we have to wait for intensive spectroscopic follow-up
observations to confirm it.
In contrast to PKS1138, we do not observe any possible (proto-)clusters
nor red sequence in the GOODS field and the galaxy density is lower.
Therefore, the two samples probe different environments at $z\sim2$ and
they provide us with a unique opportunity to quantify the dependence of
galaxy properties on density at this high redshift.

%-------------------------------

%-------------------------------

%-------------------------------
\subsection{Physical parameters from the SED fits}

\begin{figure}
\centering
\includegraphics[width=8cm]{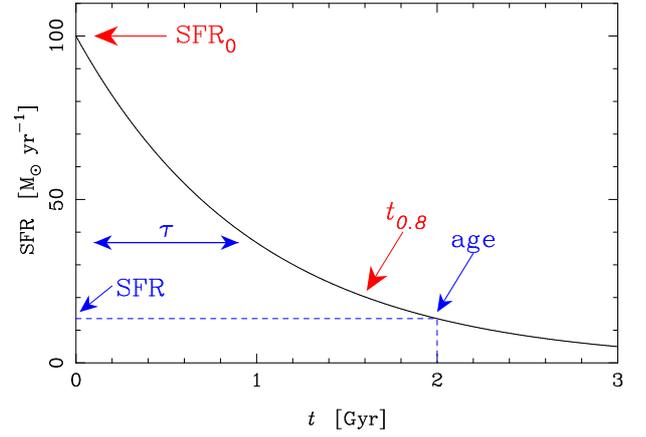}
\caption{
The parameters in the models.
We obtain star formation time scale ($\tau$), age, SFR
and dust extinction (which is not plotted) from the fits.
We then derive the initial SFR ($\rm SFR_0$) and the epoch
when galaxies form 80\% of their stars ($t_{0.8}$) and discuss them
in Section 6.
}
\label{fig:tau_model}
\end{figure}

We are now going to look into differences in galaxy
properties between the two environments.
We quantify four fundamental properties of galaxies --
age (which is time since the onset of star formation),
star formation time scale, SFR, and dust extinction
as summarized in Fig. \ref{fig:tau_model}.
We are aware of the extremely uncertain nature of various
parameters gone into the SED fits.
For example, initial mass function (IMF) is one of the most
uncertain assumptions.  A number of IMFs are suggested
in the literature, each of which produce systematically different results,
and to make matters worse, IMFs may depend on mode of star formation, 
e.g., IMFs could become top-heavy during starbursts.
Also, the $\tau$ models we adopt may not be a good approximation
of star formation histories of real galaxies, we assume all the stars
have the same metallicities, and etc.
Again, all the uncertainties are inherent in the both samples
and we discuss only the relative differences between the two samples,
which must be the most robust results from the SED fits.

%-------------------------------

We discuss the four parameters derived from the SED fits.
Since all the parameters are correlated with stellar masses of
galaxies and we plot them as a function of stellar mass.
Note that we miss a fraction of galaxies with masses lower than
$10^{11}\rm M_\odot$ in PKS1138, and
we focus on galaxies more massive than $10^{11}\rm M_\odot$.
This is a conservative mass cut and our results below
do not suffer from any strong incompleteness effects.
But, we keep lower mass galaxies in the plots below to illustrate
the mass dependency of the four parameters.
\\

%-------------------------------
\begin{figure*}[t]
\centering
\includegraphics[height=7cm]{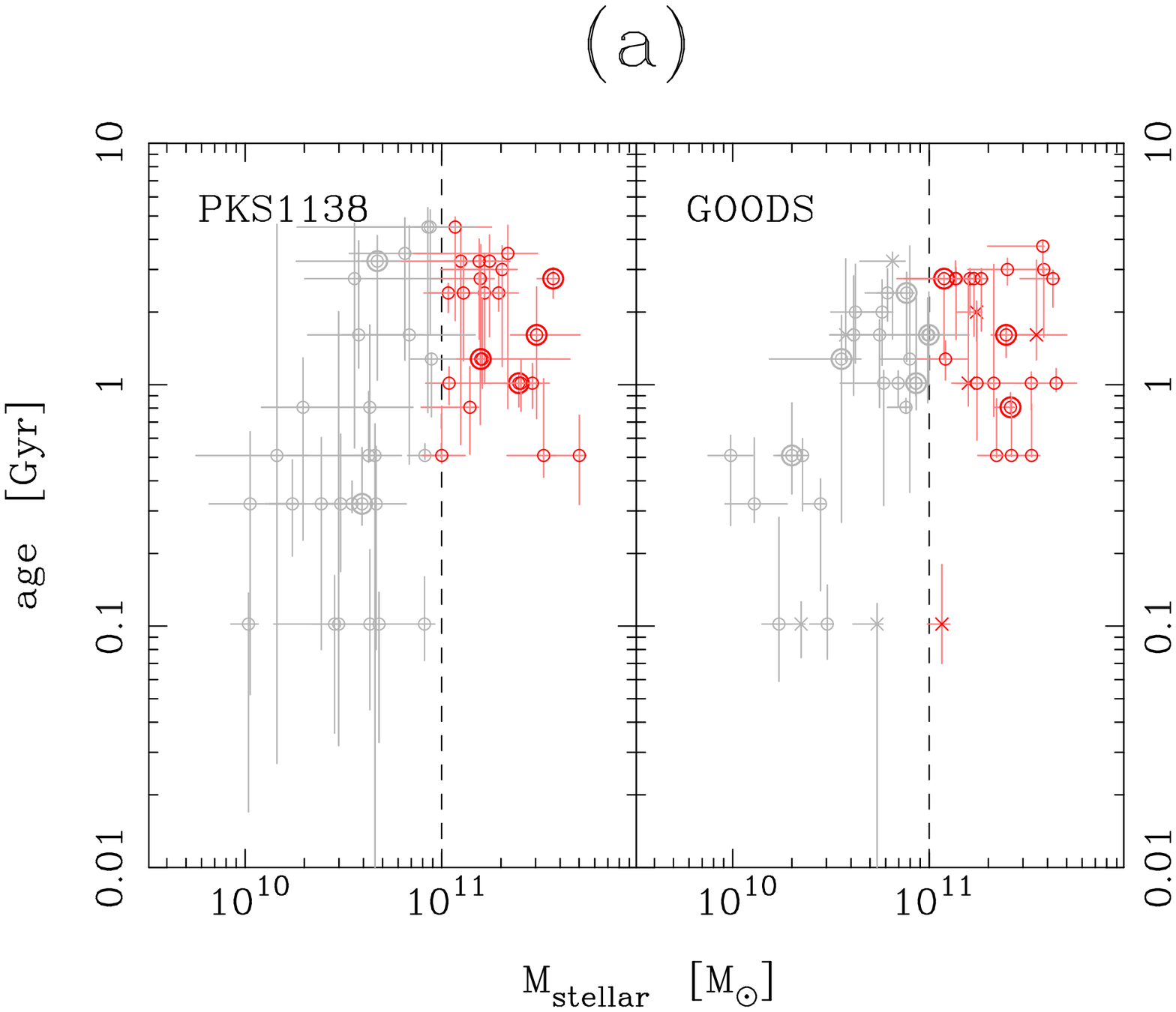}\hspace{0.5cm}
\includegraphics[height=7cm]{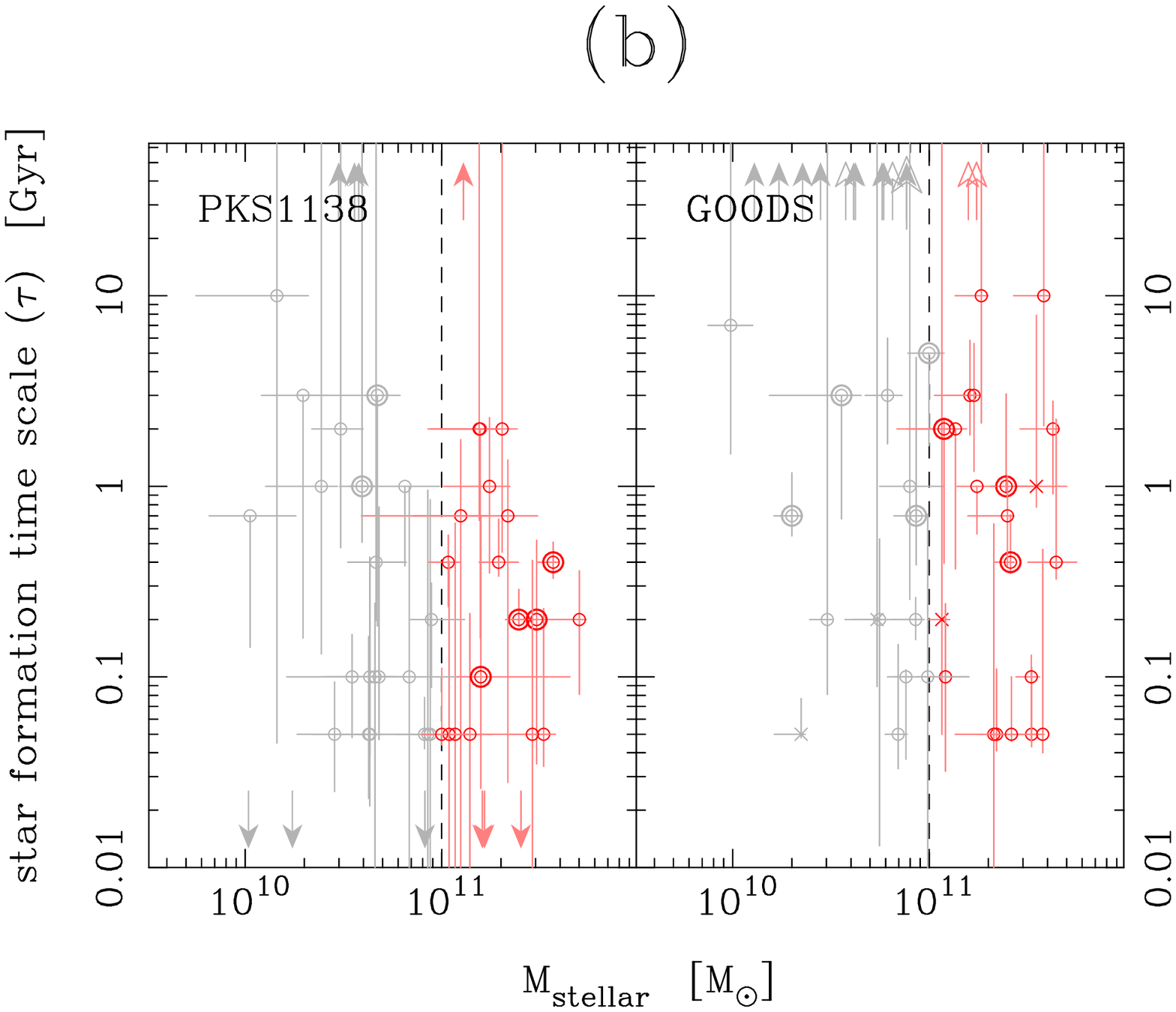}\\\vspace{0.5cm}
\includegraphics[height=7cm]{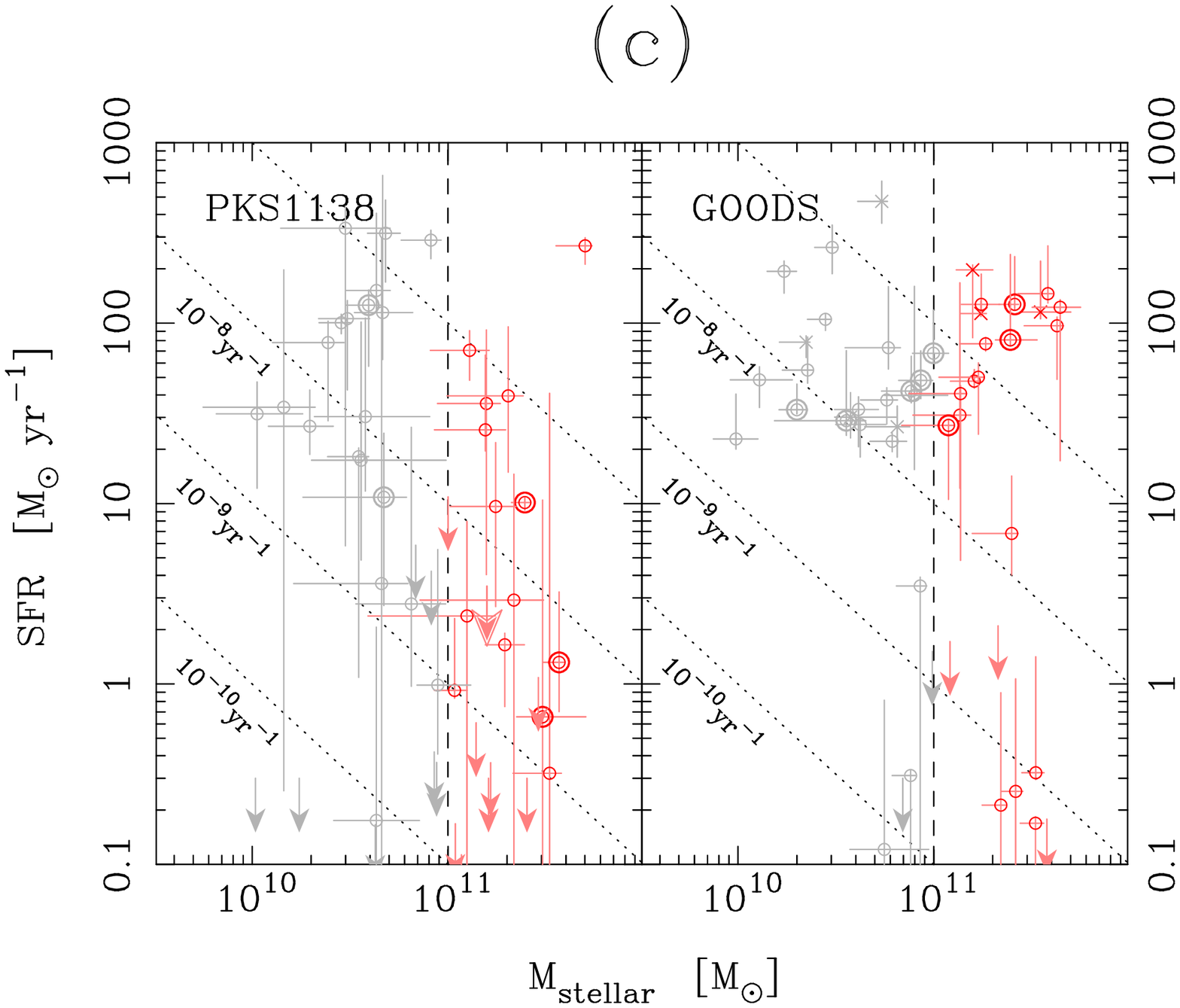}\hspace{0.5cm}
\includegraphics[height=7cm]{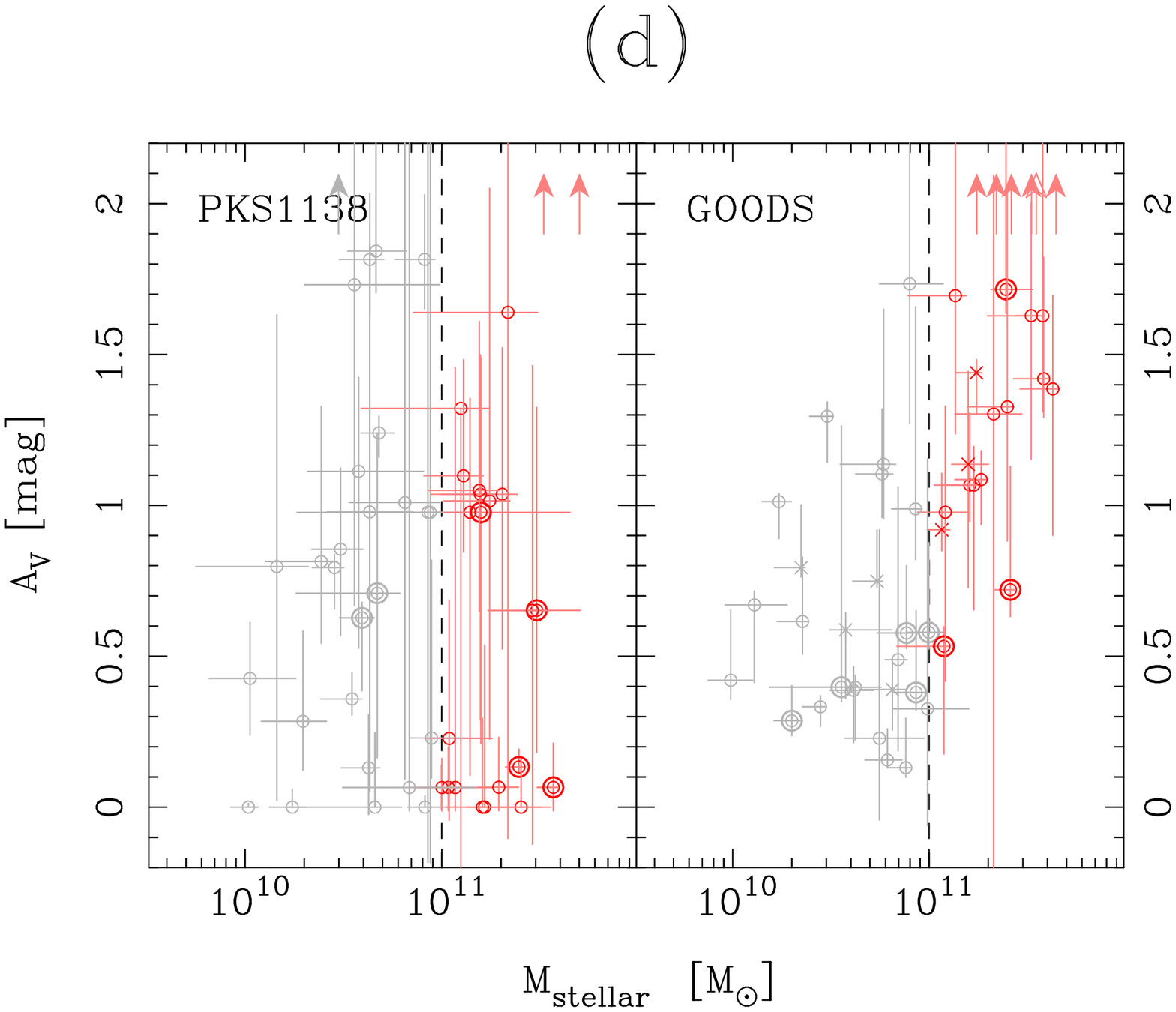}
\caption{
Physical parameters from SED fits plotted against stellar mass for PKS1138
and GOODS in the left and right panels in each plot, respectively.
The vertical dashed lines show the stellar mass limit, below which
we are incomplete.
The double-circles and crosses are spectroscopically confirmed
members and fore-/background galaxies, respectively.
For objects going outside the plotted range, we use the double arrows
and open arrows for confirmed members and outliers.
{\it (a) top-left: } Age plotted against stellar mass.
{\it (b) top-right: } Star formation time scale ($\tau$) plotted against stellar mass.
{\it (c) bottom-left: } SFR plotted against stellar mass.
The slanted dashed lines show the specific star formation rates defined by SFR / stellar mass.
{\it (d) bottom-right: } Dust extinction plotted against stellar mass.
We plot the extinction in units of $\rm A_V$.
}
\label{fig:physical_params}
\end{figure*}

\noindent
{\bf AGE (Fig. \ref{fig:physical_params}a) :}
The age distributions are not significantly different.
We find that the fractions of galaxies older than 2 Gyr are
$0.52\pm0.19$ and $0.42\pm0.16$, respectively\footnote{
%If we do not reject galaxies with poor fits ($\chi^2_\nu>3$),
As discussed in Section 4.2, we do not use galaxies with poor fits ($\chi^2_\nu>3$).
The fractions do not change if we use them; $0.59\pm0.18$ and $0.42\pm0.15$ for
PKS1138 and GOODS, respectively.
The fractions quoted below for the other parameters also remain essentially the same:
$0.72\pm0.21$ and $0.46\pm0.16$ for $\tau<$0.5 Gyr,
$0.79\pm0.22$ and $0.38\pm0.14$ for $\rm SFR<10\rm\ M_\odot\ yr^{-1}$,
and $0.66\pm0.19$ and $0.23\pm0.10$ for $\rm A_V<1$mag.
}.
Recall that the age is the time since the onset of star formation
to the observed epoch.
These face values should be taken with caution because
we did not take into account the random errors on the age estimates.
We take a Monte-Carlo approach to quantify the difference.
We let each data point fluctuate by its error and perform
the Mann-Whitney test for each realization.
We repeat the simulation for 100,000 times and find that
the median probability for the two distribution being
drawn from the same parent population is
30\%.
We obtained $<5\%$ probabilities only in 3\% of the realizations.
There is no strong evidence for age differences.

We note that \citet{steidel05} studied a spectroscopically
confirmed redshift spike at $z=2.3$ and found that
galaxies in the spike are older than those outside the spike.
Their result may appear inconsistent with ours.
But, their sample is a collection of star forming galaxies,
while we probe the entire galaxy population.
A fair comparison cannot be made.
\\

\noindent
{\bf STAR FORMATION TIME SCALE, $\tau$ (Fig. \ref{fig:physical_params}b) :} 
PKS1138 galaxies tend to have shorter formation time scales.
The fractions of galaxies with $\tau<0.5$ Gyr are
$0.70\pm0.23$ and $0.43\pm0.16$ in PKS1138 and GOODS, respectively.
The Mann-Whitney test suggests that the $\tau$ distributions
are likely different (median probability of $\sim4$\%).
We obtained $<5\%$ probabilities in 52\% of the Monte-Carlo realizations.
As seen in the GOODS data, the fore-/background contaminants do not 
favor any particular $\tau$, and the observed difference is not due
to the contamination in PKS1138.
Combining with the age distribution, most galaxies in both fields have
$\rm age > \tau$, suggesting that they already formed the bulk of
their stars by the time of the observation.
The shorter formation time scale of PKS1138 is one of the most
important results in this paper, and we will further discuss it
in the next section.\\

\noindent
{\bf SFR (Fig. \ref{fig:physical_params}c) :} 
PKS1138 galaxies tend to show weaker star formation and lower specific SFRs.
The fractions of galaxies that have $\rm SFR<10\rm\ M_\odot\ yr^{-1}$ are
$0.74\pm0.24$ and $0.33\pm0.14$ in PKS1138 and GOODS, respectively.
The Mann-Whitney test suggests that they are not from the same parent
distribution ($0.7$\%).
The fore-/background contamination tends to have high star formation.
They unlikely to contribute to the observed difference.
We recall that we obtained a lower frequency of MIPS detections
in PKS1138 than in GOODS, which gives a further support
to the lower SFRs in PKS1138.
\\

\noindent
{\bf DUST EXTINCTION  (Fig. \ref{fig:physical_params}d) :}
PKS1138 galaxies tend to have less dust, as expected from their low SFRs.
The fraction of PKS1138 galaxies with $A_V<1$ is $0.61\pm0.21$
($A_V=1$ is a typical amount of dust in local star forming galaxies).
Most of galaxies in GOODS are very dusty, and there seems a correlation
between stellar mass and dust extinction (e.g., \citealt{reddy06}).
The fraction of $A_V<1$ galaxies is $0.17\pm0.09$.
The Mann-Whitney test shows that the $A_V$ distributions
are different ($0.1$\%).
\\

To sum up, we find that galaxies in PKS1138 and GOODS have different properties.
Interestingly, the trend we observe in $z<1$ clusters still qualitatively persists
even at this high redshift --- we observe galaxies with lower SFR
and less dust in higher density regions, where we see clearer red sequence.
The environmental dependence of galaxy properties is already in place at $z=2$,
at least partly.
We will further extend the discussion and quantify the environmental
dependence of galaxy formation in the next section.

%-------------------------------------------------------
\section{Discussion}

From the extensive SED fits of galaxies in PKS1138 and GOODS,
we have found that massive galaxies in PKS1138 tend to have

\begin{enumerate}
\item similar ages
\item shorter star formation time scales
\item lower star formation rates
\item lower amounts of dust
\end{enumerate}

\noindent
compared to those in GOODS.
The combination of first and second points is interesting.
Galaxies in PKS1138 and GOODS start forming stars at a similar epoch
in a statistical sense, but PKS1138 galaxies form more rapidly.
We recall that our definition of age is the time since
the onset of star formation.
The third and fourth points are basically by-products of
the first and second points.
At $z=2$, PKS1138 galaxies have already undergone intense star formation
and their SFRs are rapidly declining, while GOODS galaxies are still
actively forming stars due to their longer star formation time scales.
This results in the lower SFRs in PKS1138 (third point).
The fourth point can then be easily understood given 
the correlation between between SFR and dust amount
(lower SFR galaxies have less dust, e.g., \citealt{hopkins03}).

Cluster galaxies have a shorter star formation time scale
-- this is the same trend as observed in $z\lesssim1$ clusters.
For example, \citet{gobat08} showed the same trend in a $z=1.2$
cluster based on a photo-spectroscopic analysis.
It is striking that the trend holds even at $z=2.15$.
The difference in the star formation time scale suggests
that cluster and field galaxies may form in different ways.
Let us discuss the formation and evolution of cluster and field galaxies
in detail.
We first introduce two parameters to further quantify the galaxy formation.
We then extend the discussion and address the origin
of the environmental dependence of galaxy properties
observed at lower redshifts.
We note a caveat here that our results are based
on one (proto-)cluster only and the trends we observe
may not represent global trends at $z=2$. A larger sample of
$z=2$ clusters should be investigated to draw a global picture.

%-------------------------------
\subsection{The environmental dependence of galaxy formation}

\begin{figure}
\centering
\includegraphics[width=8cm]{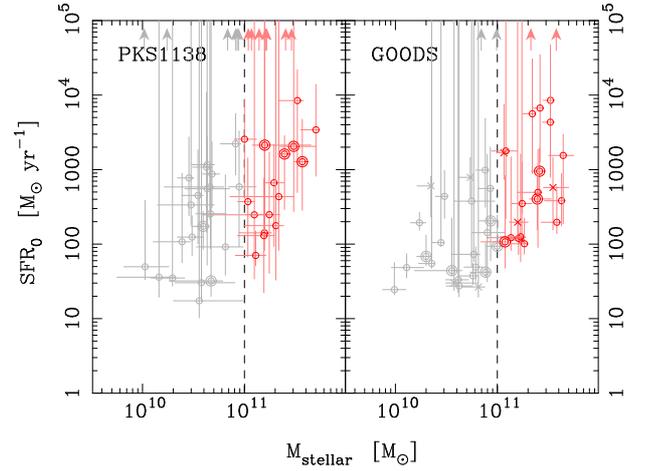}
\caption{
$SFR_0$ is plotted against stellar mass.  The meanings of the symbols
are the same as in Fig. \ref{fig:physical_params}.
}
\label{fig:sfr0_mstar}
\end{figure}

We have age, star formation time scale, and SFR for each galaxy.
Assuming the exponentially decaying star formation histories,
we can estimate the SFR at the onset of star formation,
which we call initial SFR ($\rm SFR_0$; see Fig. \ref{fig:tau_model}).
Fig. \ref{fig:sfr0_mstar} plots  $\rm SFR_0$ as a function of
stellar mass.
The cluster galaxies tend to have higher $\rm SFR_0$ --
the fractions of $\rm SFR_0>1000M_\odot\ yr^{-1}$ are $0.61\pm0.21$,
and $0.39\pm0.15$ in PKS1138 and GOODS, respectively.
The Mann-Whitney test supports this difference (the median probability is
$2$\% and we obtain $<5$\% probability in 74\% of the realizations).
PKS1138 galaxies have experienced more intense formation histories.
Of course, all this is based on the assumption of the exponentially
declining SFRs, and we have ignored effects of galaxy-galaxy mergers,
which we will discuss later.

\begin{figure}
\centering
\includegraphics[width=8cm]{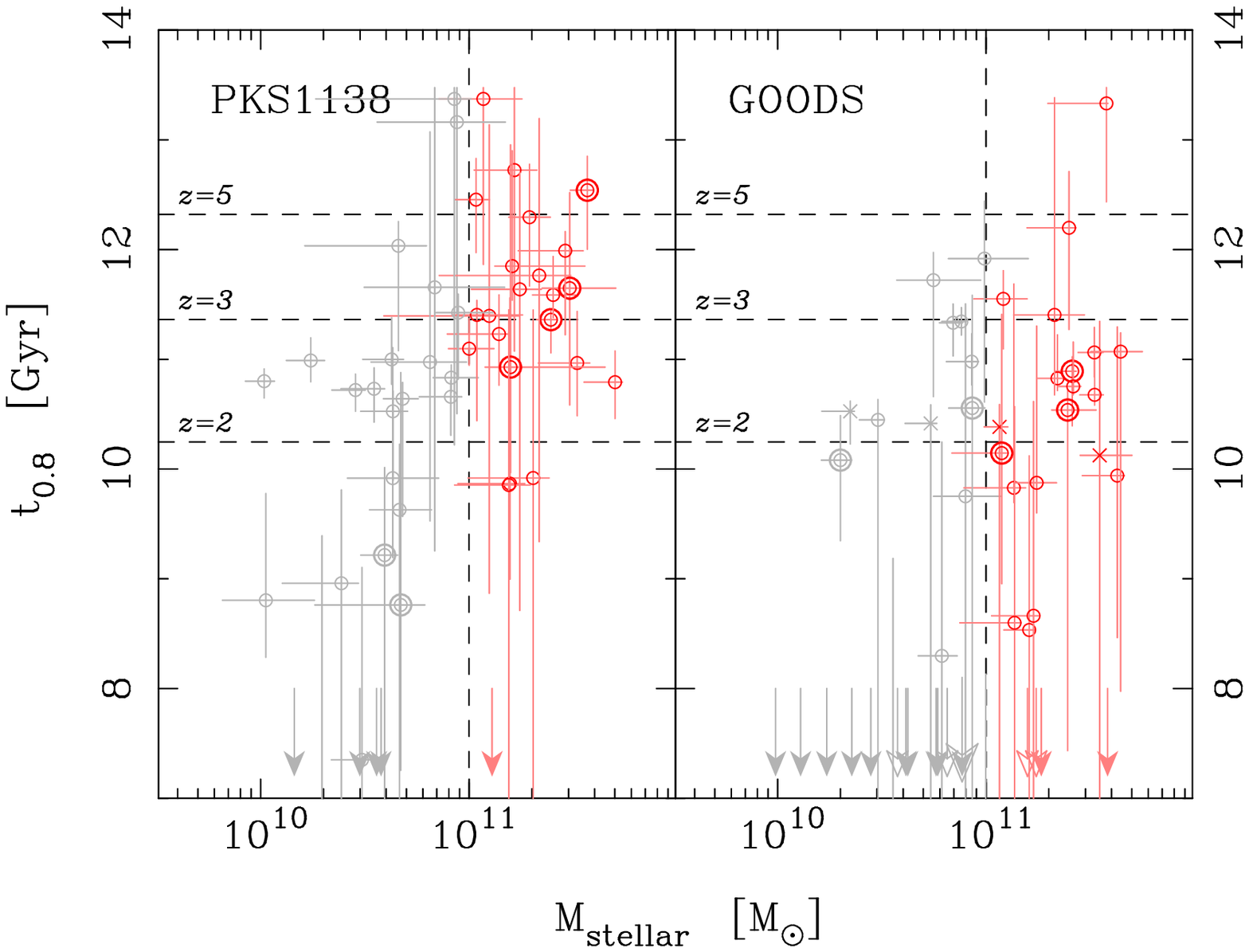}
\caption{
$t_{0.8}$ is plotted against stellar mass.  The meanings of the symbols
are the same as in Fig. \ref{fig:physical_params}.
}
\label{fig:m80_mstar}
\end{figure}

If galaxies continue to form stars following the exponential decay,
we can derive the time when the galaxies form bulk of their stars.
We plot in Fig. \ref{fig:m80_mstar} the epoch when galaxies form
80\% of the stars that they would have at $z=0$
($t_{0.8}$; see Fig. \ref{fig:tau_model}).
PKS1138 galaxies typically formed around $z\sim3$ or higher, while GOODS galaxies
typically formed below $z\sim3$.
We recall that we obtained $z_f\sim4$ from the location of the red sequence
in PKS1138 in Section 5.1, which is in agreement with what we find here.
We find that $0.61\pm0.21$ of galaxies in PKS1138 form the bulk of their stars
at $z>3$, while the fraction is $0.17\pm0.09$ in GOODS.
The Mann-Whitney test suggests that they are likely different
-- the median probability is $\sim8\%$ and we obtained 
$<5$\% probabilities in 38\% of the realization.
Interestingly, the ages of the cluster and field galaxies are not
very different (Fig. \ref{fig:physical_params}a).
The difference in $t_{0.8}$ is therefore due to their shorter formation time scales.
The formation of cluster galaxies is a more intense event, they
form in a shorter time scale, and they assemble the bulk of
their stars earlier by $\sim1-2$ Gyr than the field galaxies.
%Detailed spectral analyses of local elliptical galaxies suggest
%that cluster ellipticals form the bulk of their mass earlier
%by $\sim1$ Gyr than field ellipticals (e.g., \citealt{thomas05}).
%Our observation is fully consistent with this.

%-------------------------------

%-------------------------------

%-------------------------------
\begin{figure}
\centering
\includegraphics[width=8cm]{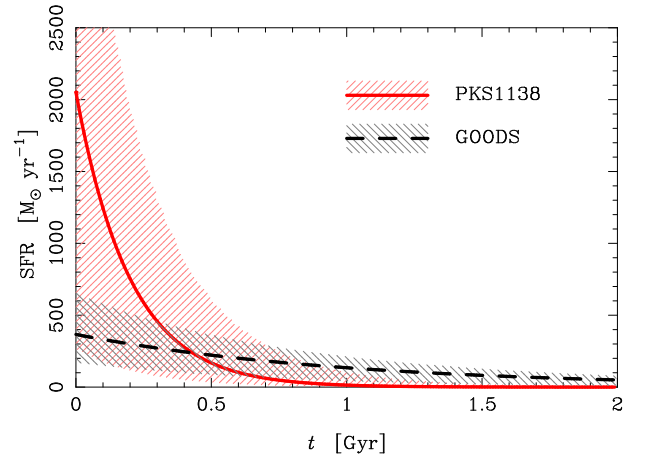}
\caption{
The median star formation histories of the PKS1138 and GOODS galaxies.
The shaded area shows 68 percentile intervals.
}
\label{fig:tau_model2}
\end{figure}

To summarize, we measure the median $\tau$ and SFR for
$>10^{11}\rm M_\odot$ galaxies
and illustrate the differences in star formation histories of galaxies
in PKS1138 and GOODS in Fig. \ref{fig:tau_model2}.
This will the summary plot of the paper.
As discussed above, PKS1138 galaxies have higher $\rm SFR_0$ and shorter
$\tau$. 
PKS1138 experience much more intense galaxy formation at early times
than GOODS galaxies.  The galaxy assemblies are completed
in a short time, and by the time we observe them ($\sim2$ Gyr after
the onset of star formation as shown in Fig. \ref{fig:physical_params}a)
their SFRs become low.
The plot summarizes the points listed at the beginning of this section
(except for the fourth one, which is a result of SFR-dust correlation).
%Interestingly, this trend is similar to the trend observed in $z\sim1$ clusters
%(e.g., \citealt{gobat08}).
%In other words, the trend seen at $z\sim1$ still persists at $z\sim2$.

The plot indicates that PKS1138 galaxies and GOODS galaxies did not form
in the same way.  The formation of cluster galaxies is a more intense event.
But, how can massive galaxies form on such a short time scale?
In general, star formation activities have negative feedback to themselves.
Hot, young stars ionize surrounding gas, which prevents further star formation.
Also, supernova explosions of massive stars give negative feedback.
If galaxies undergo an AGN phase, it might also give negative feedback.
The formation of massive galaxies on a very short time scale may be difficult
to understand in this respect, and that makes us speculate that
these massive galaxies might have taken another route -- mergers.

Galaxy formation takes place in high peaks of density fluctuations in the Universe.
Clusters that we observe today were regions of large-scale over-densities,
where many density peaks were embedded.
On the other hand, field galaxies form in more isolated peaks.
The current galaxy formation theories predict that small galaxies form first
and they form progressively massive galaxies via mergers.
In the early Universe, cluster galaxies must have had a lot better chances to
merge with other galaxies than field galaxies simply because there are
more galaxies around them.
This accelerated frequency of mergers might be the cause of the rapid
formation of cluster galaxies.

A simple scenario of an equal mass merger can be helpful here to
show that mergers tend to make $\rm SFR_0$ higher and push $t_{0.8}$
to higher redshifts.
Assuming that two equal mass galaxies merge into a single galaxy,
the merger event will double the stellar mass of the galaxy.
If the galaxy experiences starburts triggered by the merger, its stellar
mass will be more than double and its SFR gets lower after the burst.
We then observe this massive galaxy with low SFR, and depending on
the time elapsed after the burst, we would fit a short $\tau$
to this galaxy in order to reproduce its low SFR and large stellar mass.
%As discussed above, short $\tau$'s are the primary driver of
%the observed differences between PKS1138 and GOODS.
In fact, we find that the secondary burst discussed in Section 4.4
reduces the median $SFR_0$ of the PKS1138 galaxis by $\sim30\%$, although
it is still high, $SFR_0=1600\rm M_\odot\ yr^{-1}$, which
might indicate that galaxies experienced more than one merger.
On the other hand, the $SFR_0$ of the GOODS galaxies remain the same
within $10\%$.

We are not yet sure if PKS1138 is a collapsed cluster or currently
collapsing proto-cluster.  But, even if it is a cluster, it must be
a young system, and it is unlikely that galaxies had enough time
to be affected by 'nurture' effects such as ram-pressure stripping.
We may suggest that mergers in the early times of the cluster
formation may be one of the key processes to establish the environmental
dependence of galaxy properties observed at lower redshifts.
We will further pursue this point in the following section.

%-------------------------------
\subsection{The Nature effects in the environmental dependence of galaxy properties}

Galaxy properties such as colors and morphology are known to depend
on environment in which galaxies reside.
This environmental dependence is shaped by two effects -- nature and nurture effects.
That is, both how galaxies form and how they evolve are important.

Fig. \ref{fig:m80_mstar} shows that only $\sim1$ Gyr has passed
since the PKS1138 galaxies formed bulk of their stars to the observed epoch.
Assuming that PKS1138 is a virialized system, we apply a $2\sigma$-clipped
gapper method \citep{beers90} and obtain a velocity dispersion
of 400 $\rm km\ s^{-1}$ using spectroscopic redshifts within 2 arcmin,
corresponding to a physical scale of 1 Mpc, from the literature
\citep{pentericci00,kurk04a,croft05}.
The virial radius ($r_{200}$) of the cluster is 0.32 Mpc, giving a crossing
time scale of $\sim 1$ Gyr.
Therefore, the time elapsed since the bulk formation of the PKS1138 galaxies
is comparable to the crossing time scale of this cluster.
This suggests that nurture effects may not have had enough time
to fully work.
They may have affected a fraction of galaxies, but they probably
could not change the average properties of galaxies.
We may be witnessing the nature effects in shaping
the environmental dependence in PKS1138.

We admit that it is not very straightforward to classify
nature and nurture effects at this high redshift.
Now, let us introduce two kinds of mergers; early-epoch mergers
and late-epoch mergers.
We refer to mergers occur during the first collapse of clusters
as early-epoch mergers, and those occur afterwards as late-epoch mergers.
This is to sort out the two effects, early-epoch mergers being
nature effects (initial conditions) and late-epoch mergers being
nurture effects (environmental effects).
Of course, nurture effects also include ram-pressure stripping, harassment, etc.
On the other hand, nature effects do not include these processes
driven by the deep potential well or intracluster medium.

In this classification, the environmental dependence of PKS1138
is likely due to early-epoch mergers
because $\sim1$ Gyr time is probably not enough for nurture effects to fully work.
Galaxy clusters at $z\sim1$ do not give useful information
about nature effects because
nurture effects have had enough time to fully operate
($\sim4$ Gyr since the formation epoch) and
it is not straightforward to disentangle the two effects.
Our results suggest that nature effects are strong and they form
the basis of the environmental dependence of galaxy properties.
A way to probe the significance of the nature effects is to
quantify morphology of the PKS1138 galaxies.
If early-epoch mergers are an important effect, then we expect to
observe more signs of recent interactions in PKS1138 than in GOODS (see below).
Also, we expect to observe post-starburst galaxies if galaxies
undergone interaction triggered starbursts.
A near-IR spectroscopic follow-up campaign of PKS1138 is currently underway.
We will be able to study spectral properties of the PKS1138 galaxies,
which allow us to look deeper into star formation histories.

To sum up, the strong nature effects may shape
the environmental dependence of galaxy properties.
An ultimate goal of environment studies will be to quantify
the relative contribution of nature and nurture effects.
But, that will require statistical work on $z\sim2$ (proto-)clusters.
Although the possible significance of nature effects we suggest here
is based only on one (proto-)cluster PKS1138, let us further discuss it.
It has an interesting implication for the build-up of the cluster red sequence.

%-------------------------------
\subsection{The massive end of the cluster red sequence}

The sequence of red early-type galaxies is a ubiquitous feature of
galaxy clusters.
Over the last few years, there is an accumulating amount of
evidence in the literature that the cluster red sequence grows
from the massive end to the low-mass end
(e.g., \citealt{tanaka05,tanaka07a,koyama07,tanaka08,gilbank08}).
The massive end of the red sequence cannot be formed via
a simple fading of blue, star forming galaxies because
such massive blue galaxies do not exist even at $z\sim2$
(see Fig. \ref{fig:cmd} and discussions in \citealt{faber07}).
We need mergers to form it.

An interesting point here is that we observe the brightest tip
of the red sequence in PKS1138, which was also noted by \citet{zirm08}.
The massive end of the red sequence in a young system --
this might be due to early-epoch mergers.
Early-epoch mergers might have formed very massive galaxies
during the first gravitational collapse of clusters.
Early-epoch mergers should occur more frequently in
cluster environments than in the field, and that
helps explain why we do not observe red sequence in GOODS.
As suggested by \citet{zirm08}, the red sequence in PKS1138 may be
being formed or just formed at the time of observation.
This view is further supported by our observation that
roughly half of the red sequence galaxies have
high SFRs (Fig. \ref{fig:physical_params}).
This formation redshift of $z\sim2$ is in line
with predictions from the build-up of the red sequence
observed in lower redshift clusters \citep{tanaka07a}.

\begin{figure*}[tbh]
\centering\noindent
\includegraphics[width=4.5cm]{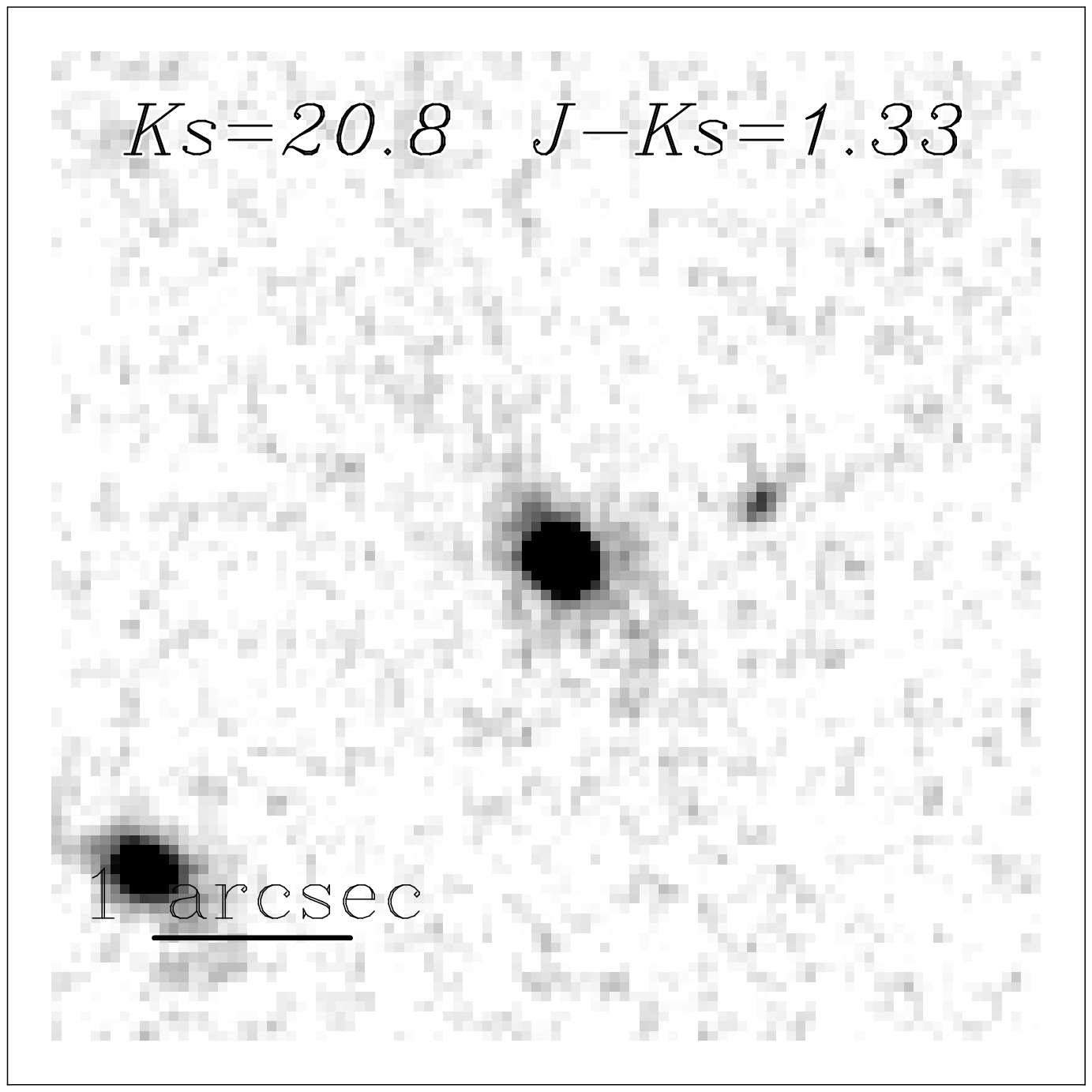}
\includegraphics[width=4.5cm]{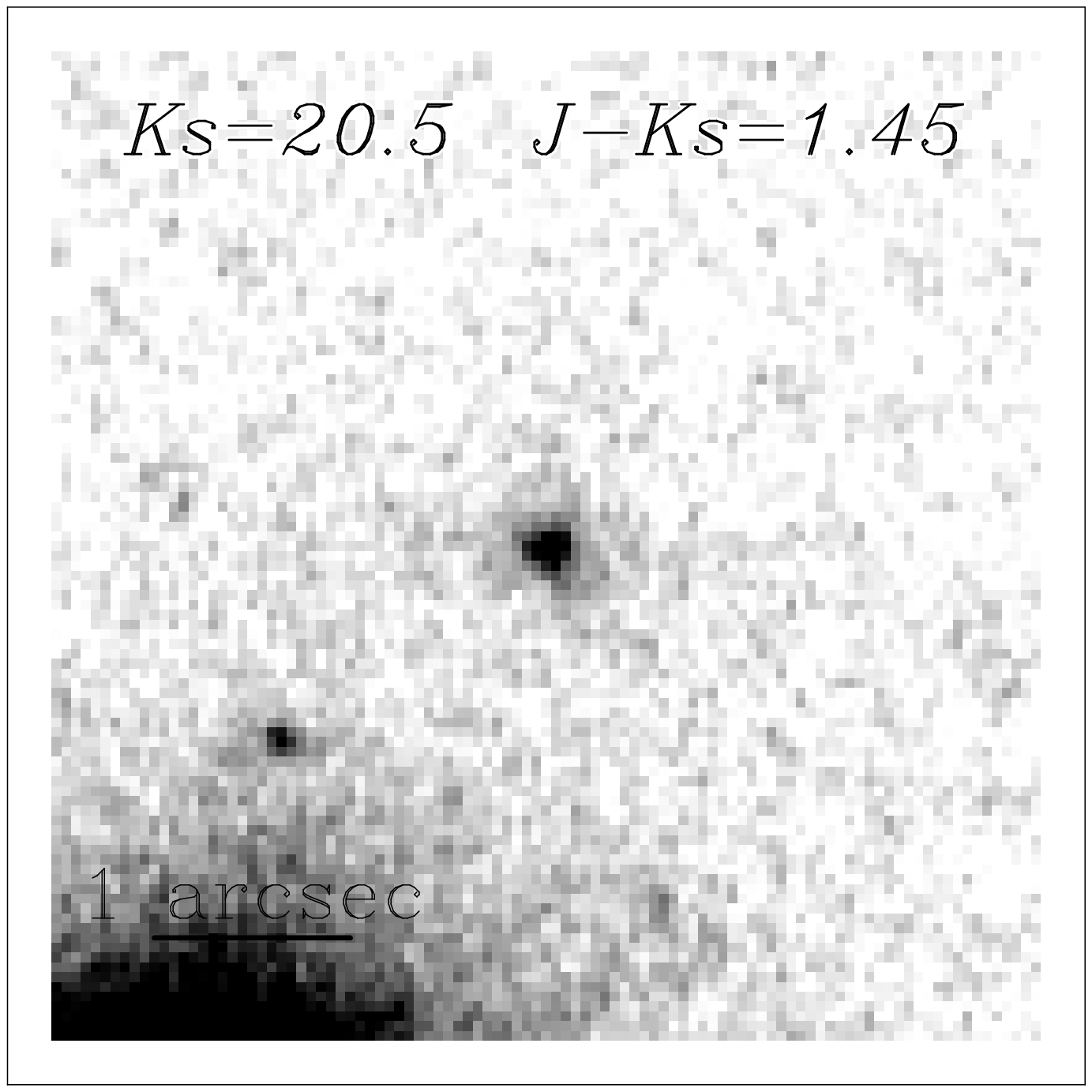}
\includegraphics[width=4.5cm]{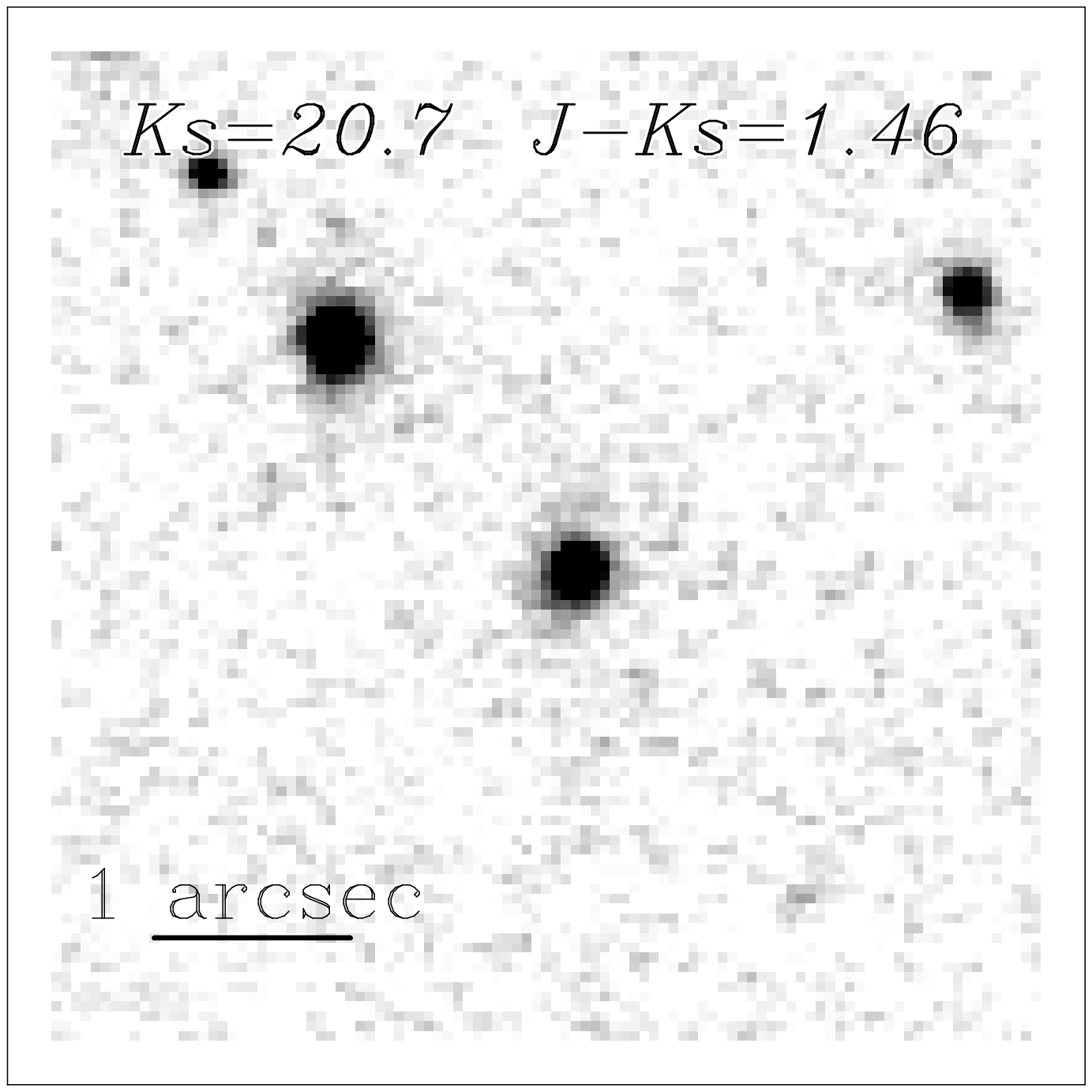}\\
\includegraphics[width=4.5cm]{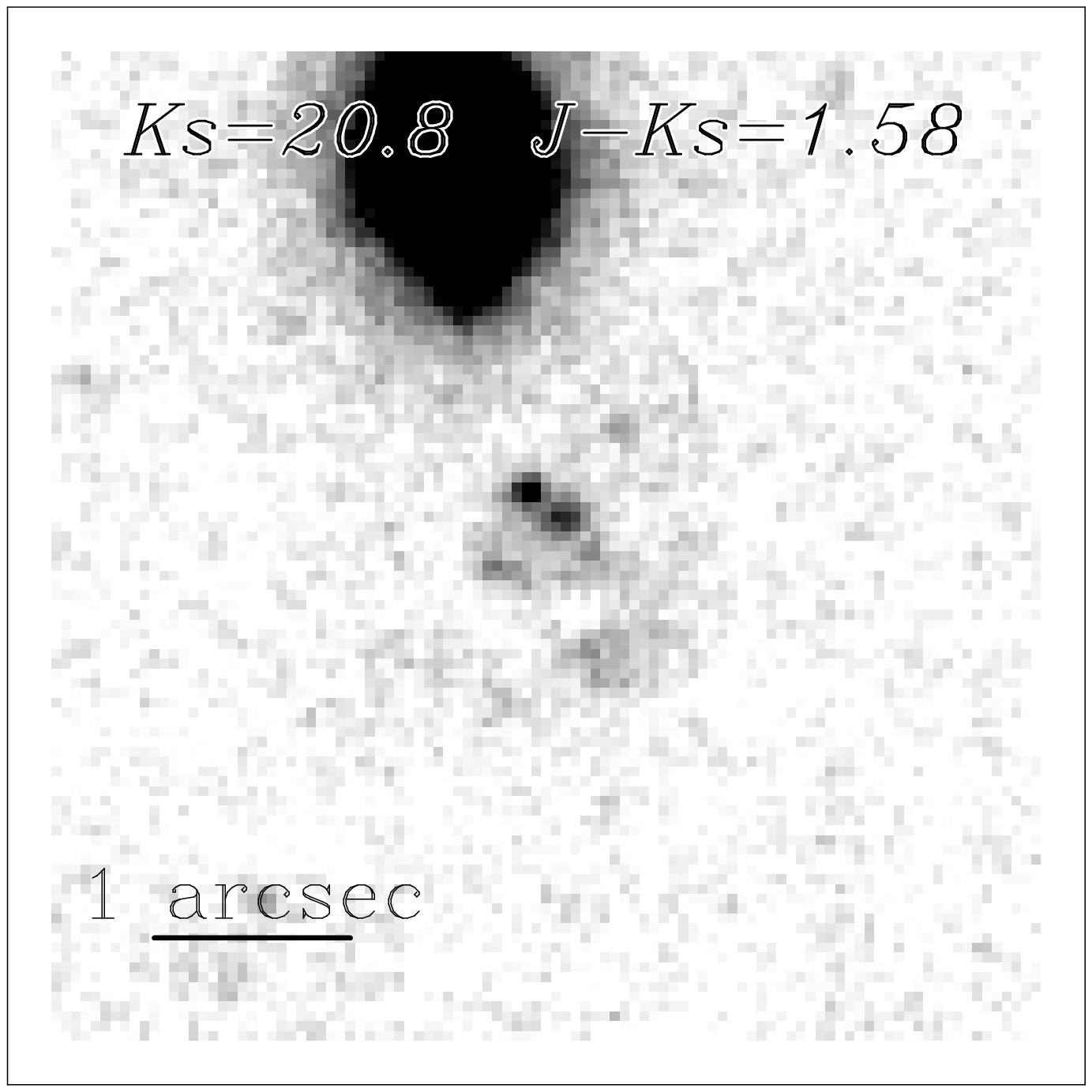}
\includegraphics[width=4.5cm]{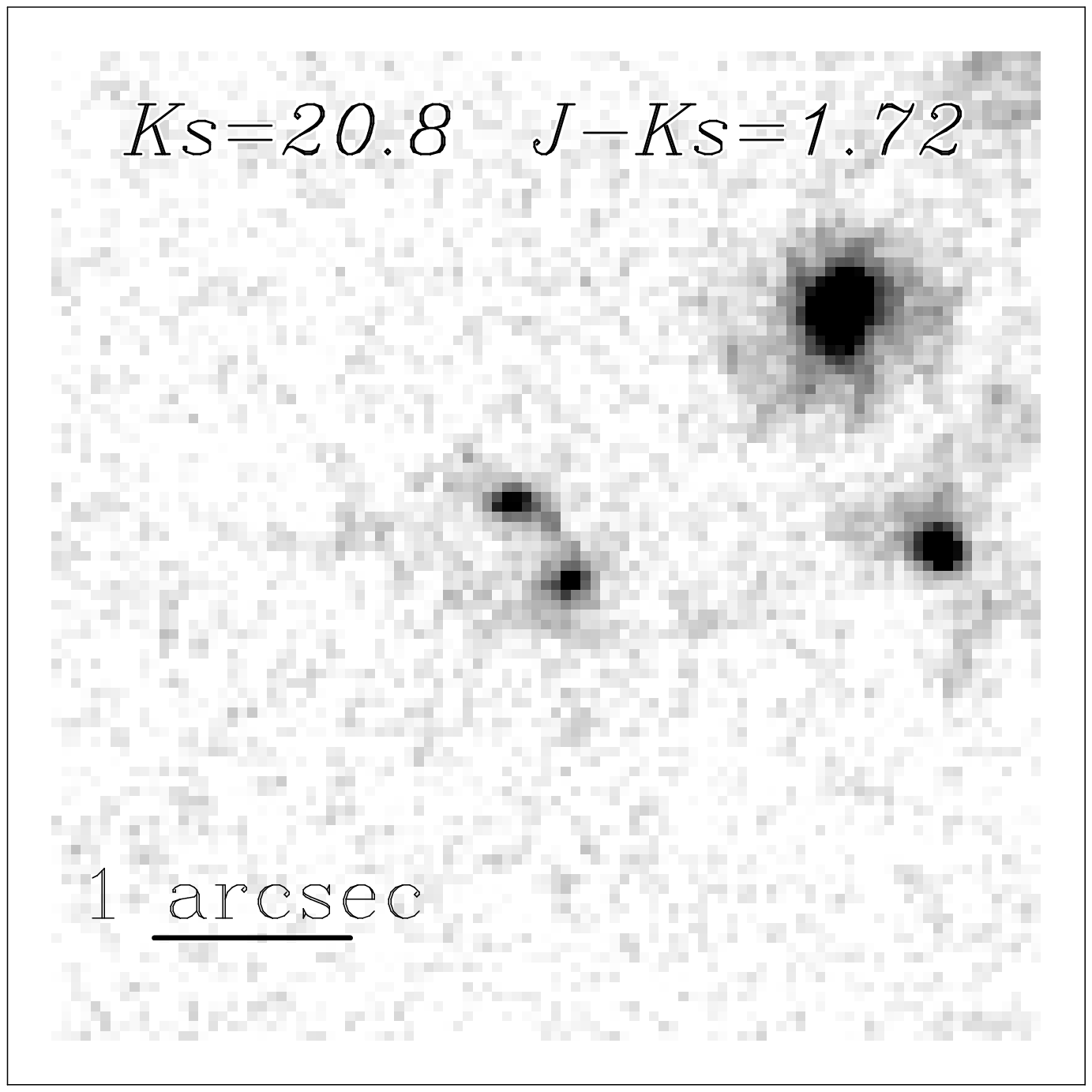}
\includegraphics[width=4.5cm]{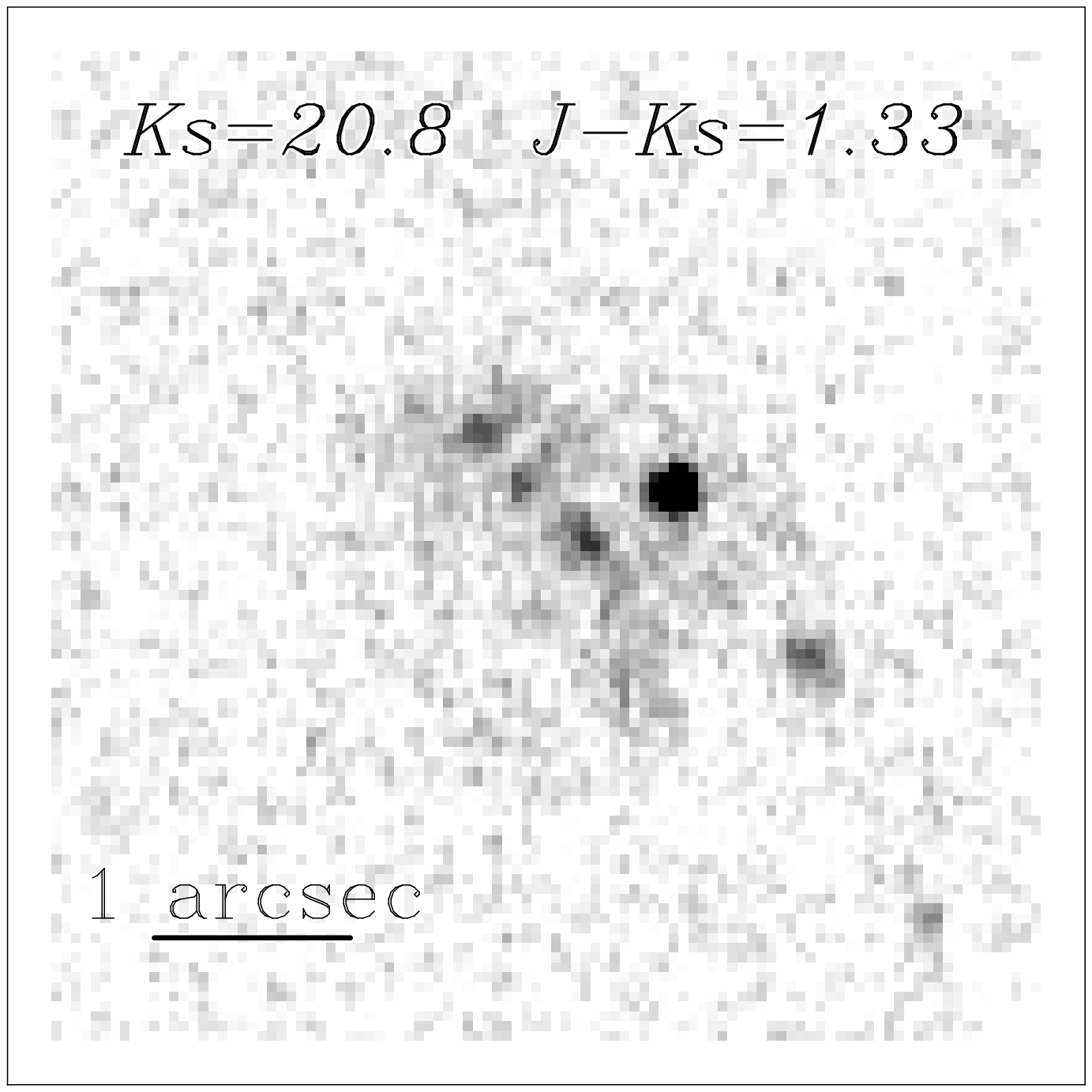}
\caption{
ACS $I$-band images of the bright red galaxies.
}
\label{fig:acs_blowup}
\end{figure*}

We present in Fig. \ref{fig:acs_blowup} ACS $I$-band images of
the bright red galaxies in PKS1138 ($K_s<21$ and $J-Ks>1$).
Here we only briefly discuss their morphologies and a detailed
study will be presented elsewhere (Zirm et al. in prep).
Interestingly, a half of the galaxies show disturbed morphology
and/or have nearby companions, lending a support to the picture
of the accelerated mergers in clusters.
Two out of three apparently disturbed galaxies are detected in MIPS.
Early-epoch mergers during the first collapse of clusters
may form the brightest end of the red sequence, and at the same time,
form the basis of the environmental dependence of galaxy properties.

%-------------------------------

The low-mass end of the red sequence is built up at later times.
Nurture effects may come in there.
Once clusters form, intracluster gas and deep potential field,
in addition to late-epoch mergers,
can affect galaxies and terminate their star formation activities.
In this way, moderate-low mass galaxies could become red and
form the low-mass end of the red sequence.
The combination of nature and nurture effects may have
conspired to produce the down-sizing behavior of the observed build-up of
the red sequence (e.g., \citealt{tanaka05}).

%-------------------------------
\subsection{Proto-clusters at higher redshifts}

Finally, we finish the discussion with a forecast for
future (proto-)cluster studies.
We can go back in time and see how the PKS1138 field looked
like at higher redshifts as we have star formation histories
of individual galaxies.
A caveat of course is that we ignore all the early-epoch mergers
occurred before the time of the observation (the only way to
recover that information is to resolve the galaxies
into individual stars).
Another caveat is that we cannot track spatial positions of
galaxies back in time.  But, PKS1138 is likely a collapsing/collapsed
system and it may well have been an over-density region already at $z=4$.

\begin{figure*}
\centering
\includegraphics[height=10cm]{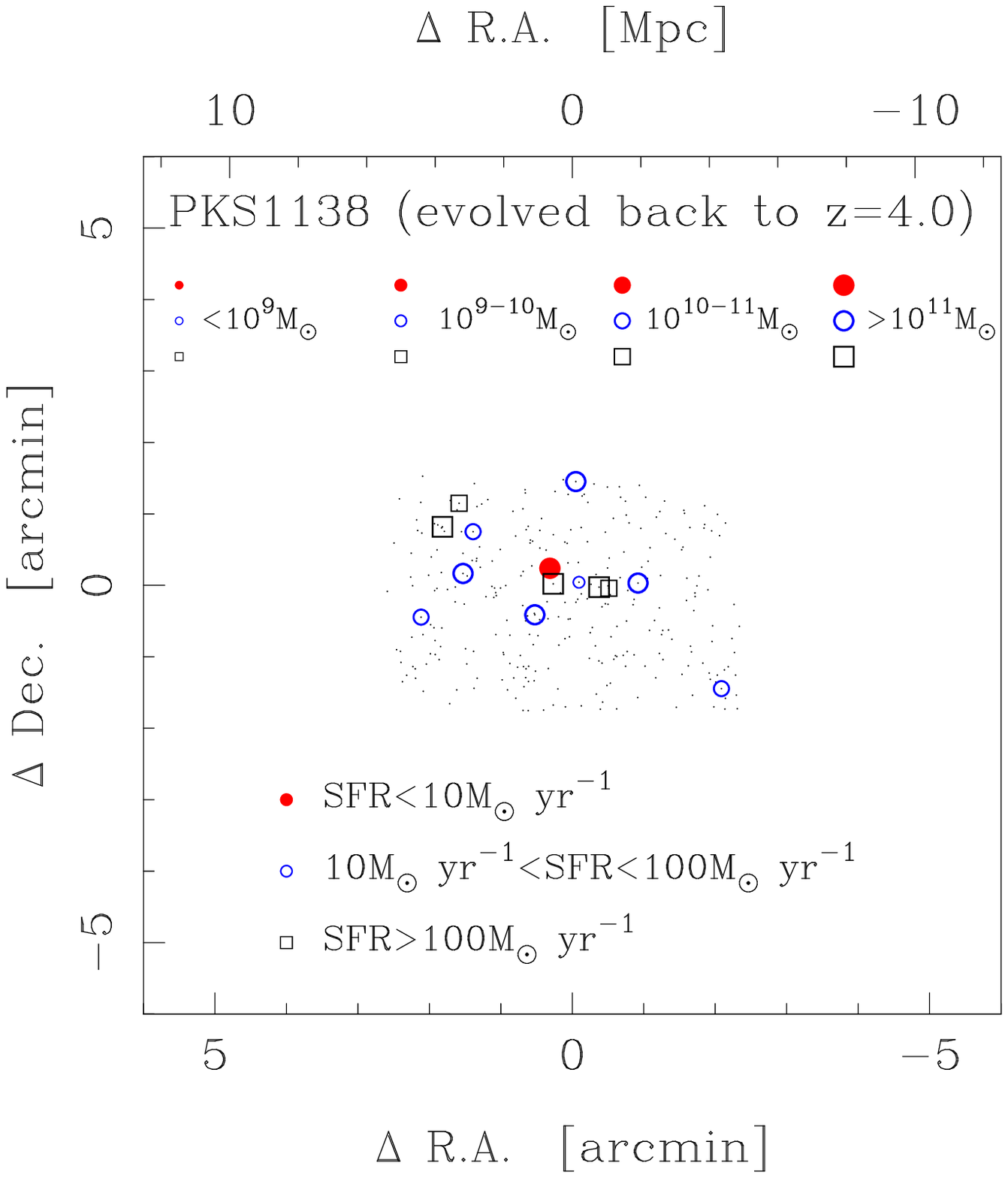}
\includegraphics[height=10cm]{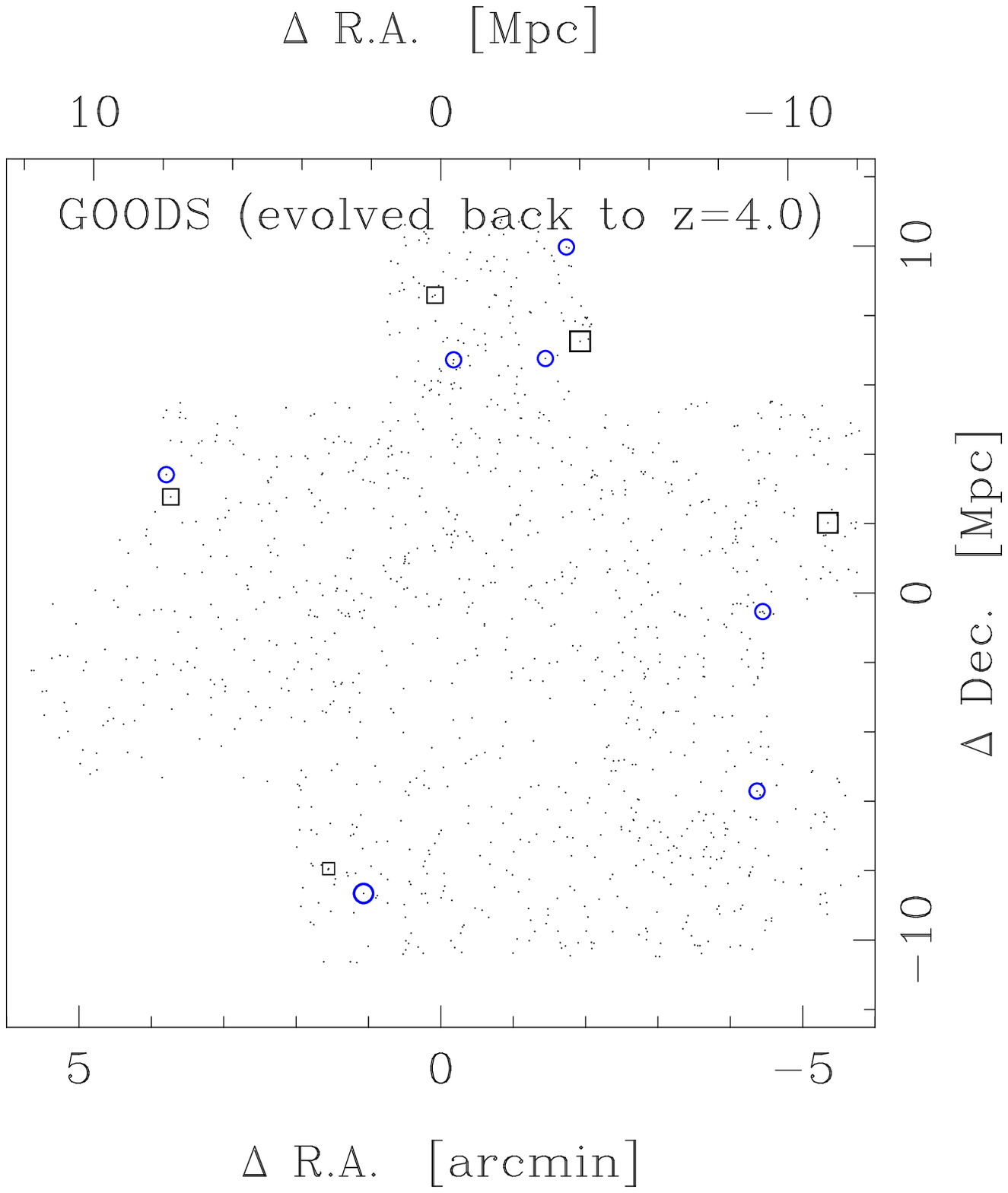}
\caption{
Same as Fig. \ref{fig:lss}, but galaxies are evolved back in time to
$z=4$ assuming the star formation histories obtained from the SED fits.
The sizes of the symbols correlate with stellar mass.
The top and right axes show comoving scales at $z=4$.
}
\label{fig:lss_evol}
\end{figure*}

Fig. \ref{fig:lss_evol} shows distribution of galaxies
and their star formation rates evolved back to $z=4$.
Compared to Fig. \ref{fig:lss}, galaxies are more actively forming
stars on average.  Interestingly, galaxies in PKS1138 are
more actively forming stars and there are more starbursting
galaxies with $\rm SFR>100\ M_\odot\ yr^{-1}$ than in GOODS.
This is in stark contrast to Fig. \ref{fig:lss}, where we saw
that PKS1138 galaxies have lower SFRs than those in GOODS.
We still observe a hint of a galaxy over density in PKS1138 at $z=4$.
The plots suggest that, as we approach the formation
epoch of clusters (i.e., early phase of the gravitational collapse to
a massive cluster halo), we expect to observe an over density of
starbursting galaxies.
Along with starbursting galaxies, low SFR galaxies already
appear in PKS1138, while such galaxies are extremely rare in GOODS.
At this redshift, low SFR galaxies and starbursting galaxies
may co-exist in a forming cluster.

It is not easy to show how early-epoch mergers
change the picture we see here.  But, we still expect to
observe starbursting galaxies triggered by early-epoch
mergers in collapsing clusters with a higher over-densities of
lower-mass, pre-merger galaxies.
We deem that observations with existing/future
sub-millimeter arrays would be able to discover many
forming clusters at high redshifts.
In fact, some submm observations of distant radio
galaxies have found possible over-densities of dusty
starburst populations around them (e.g., \citealt{debreuck04,greve07}).
Full wavelength observations will be essential for
future proto-cluster studies at very high redshifts.

%-------------------------------

%-------------------------------------------------------
\section{Summary}

We have studied the environmental dependence of galaxy properties
at $z\sim2$ based on the multi-band data available in
the (proto-)cluster field PKS1138 and in GOODS.
We have performed the extensive SED fits with a special care for
systematic biases between the two samples.
The results from the careful SED fits suggest that
the environmental dependence is at least partly in place
at this high redshift.

We have first shown that PKS1138 is indeed an over-density region
with an excess of red galaxies, forming the brightest tip of
the red sequence, compared to GOODS.
The red galaxies tend to cluster around the radio galaxy.
These results support the claim that PKS1138 is
a (proto-)cluster at $z=2.15$.
Interestingly, the red sequence is populated both by
low SFR galaxies and high SFR galaxies, suggesting
that the red sequence is being formed.

We then have looked into detailed properties of galaxies
derived from the SED fits.  PKS1138 galaxies have
similar age (we define age as time since the onset of
star formation), shorter star formation time scale,
lower SFR, and less dust compared to those in GOODS
at similar redshifts.
The averaged star formation history of the PKS1138
suggests that the cluster galaxies form on a shorter
time scale and they form the bulk of their stars
$\sim1$ Gyr earlier than the field galaxies, which is
consistent with lower redshift observations
(e.g., \citealt{thomas05,gobat08}).

The environmental dependence of galaxy properties should be
shaped both by nature and nurture effects.
The observed environmental dependence at $z=2.15$ suggests
that nature effects may be a strong effect as the PKS1138
(proto-)cluster is likely a young system.
Possibly an accelerated  rate of
mergers in collapsing clusters may have played
a role there and they may also be the primary cause of
the short formation time scale of cluster galaxies.
However, further studies on $z>2$ (proto-)clusters will
be needed to confirm this picture.
We expect that star bursting galaxies populate in very high
redshift proto-clusters and sub-millimeter observations
may be useful to confirm very high-$z$ systems.

%-------------------------------------------------------
\begin{acknowledgements}
We thank Nick Seymour, Tadayuki Kodama and Andrew Zirm for providing us
with their data on PKS1138, Chris Lidman for reducing the SOFI data
obtained at the European Southern Observatory using the Very Large Telescope
on Cerro Paranal through ESO program 66.A-0597 and 167.A-0409, and
also using the New Technology Telescope on Cerro La Silla through ESO program
076.A-0670.
We thank Ricardo Demarco for helping with the NTT observation, and 
Stephane Charlot and Gustavo Bruzual for providing us with
their latest stellar population synthesis code.
We are grateful to the anonymous referee for useful comments,
which helped improve the paper.
This work was supported by World Premier International Research Center 
Initiative (WPI Initiative), MEXT, Japan.
JK thanks the DFG for support via German-Israeli Project Cooperation grant
STE1869/1-1.GE625/15-1.
This study is also based on observations made with the NASA/ESA Hubble Space Telescope,
and obtained from the Hubble Legacy Archive, which is a collaboration between
the Space Telescope Science Institute (STScI/NASA), the Space Telescope
European Coordinating Facility (ST-ECF/ESA) and the Canadian Astronomy
Data Centre (CADC/NRC/CSA). 
\end{acknowledgements}

%-------------------------------------------------------
\bibliographystyle{aa}
\bibliography{1138_sedfit_refs}

\begin{thebibliography}{62}
\expandafter\ifx\csname natexlab\endcsname\relax\def\natexlab#1{#1}\fi

\bibitem[{{Beers} {et~al.}(1990){Beers}, {Flynn}, \& {Gebhardt}}]{beers90}
{Beers}, T.~C., {Flynn}, K., \& {Gebhardt}, K. 1990, \aj, 100, 32

\bibitem[{{Bertin} \& {Arnouts}(1996)}]{bertin96}
{Bertin}, E. \& {Arnouts}, S. 1996, \aaps, 117, 393

\bibitem[{{Blakeslee} {et~al.}(2003){Blakeslee}, {Franx}, {Postman}, {Rosati},
  {Holden}, {Illingworth}, {Ford}, {Cross}, {Gronwall}, {Ben{\'{\i}}tez},
  {Bouwens}, {Broadhurst}, {Clampin}, {Demarco}, {Golimowski}, {Hartig},
  {Infante}, {Martel}, {Miley}, {Menanteau}, {Meurer}, {Sirianni}, \&
  {White}}]{blakeslee03}
{Blakeslee}, J.~P., {Franx}, M., {Postman}, M., {et~al.} 2003, \apjl, 596, L143

\bibitem[{{Bruzual} \& {Charlot}(2003)}]{bruzual03}
{Bruzual}, G. \& {Charlot}, S. 2003, \mnras, 344, 1000

\bibitem[{{Chabrier}(2003)}]{chabrier03}
{Chabrier}, G. 2003, \pasp, 115, 763

\bibitem[{{Charlot} \& {Fall}(2000)}]{charlot00}
{Charlot}, S. \& {Fall}, S.~M. 2000, \apj, 539, 718

\bibitem[{{Cooper} {et~al.}(2008){Cooper}, {Newman}, {Weiner}, {Yan},
  {Willmer}, {Bundy}, {Coil}, {Conselice}, {Davis}, {Faber}, {Gerke},
  {Guhathakurta}, {Koo}, \& {Noeske}}]{cooper08}
{Cooper}, M.~C., {Newman}, J.~A., {Weiner}, B.~J., {et~al.} 2008, \mnras, 383,
  1058

\bibitem[{{Croft} {et~al.}(2005){Croft}, {Kurk}, {van Breugel}, {Stanford}, {de
  Vries}, {Pentericci}, \& {R{\"o}ttgering}}]{croft05}
{Croft}, S., {Kurk}, J., {van Breugel}, W., {et~al.} 2005, \aj, 130, 867

\bibitem[{{Cucciati} {et~al.}(2006){Cucciati}, {Iovino}, {Marinoni}, {Ilbert},
  {Bardelli}, {Franzetti}, {Le F{\`e}vre}, {Pollo}, {Zamorani}, {Cappi},
  {Guzzo}, {McCracken}, {Meneux}, {Scaramella}, {Scodeggio}, {Tresse}, {Zucca},
  {Bottini}, {Garilli}, {Le Brun}, {Maccagni}, {Picat}, {Vettolani},
  {Zanichelli}, {Adami}, {Arnaboldi}, {Arnouts}, {Bolzonella}, {Charlot},
  {Ciliegi}, {Contini}, {Foucaud}, {Gavignaud}, {Marano}, {Mazure}, {Merighi},
  {Paltani}, {Pell{\`o}}, {Pozzetti}, {Radovich}, {Bondi}, {Bongiorno},
  {Busarello}, {de La Torre}, {Gregorini}, {Lamareille}, {Mathez}, {Mellier},
  {Merluzzi}, {Ripepi}, {Rizzo}, {Temporin}, \& {Vergani}}]{cucciati06}
{Cucciati}, O., {Iovino}, A., {Marinoni}, C., {et~al.} 2006, \aap, 458, 39

\bibitem[{{De Breuck} {et~al.}(2004){De Breuck}, {Bertoldi}, {Carilli},
  {Omont}, {Venemans}, {R{\"o}ttgering}, {Overzier}, {Reuland}, {Miley},
  {Ivison}, \& {van Breugel}}]{debreuck04}
{De Breuck}, C., {Bertoldi}, F., {Carilli}, C., {et~al.} 2004, \aap, 424, 1

\bibitem[{{Doherty et al.}(2009)}]{doherty09}
{Doherty et al.} 2009, \aap, submitted

\bibitem[{{Elbaz} {et~al.}(2002){Elbaz}, {Cesarsky}, {Chanial}, {Aussel},
  {Franceschini}, {Fadda}, \& {Chary}}]{elbaz02}
{Elbaz}, D., {Cesarsky}, C.~J., {Chanial}, P., {et~al.} 2002, \aap, 384, 848

\bibitem[{{Elbaz} {et~al.}(2007){Elbaz}, {Daddi}, {Le Borgne}, {Dickinson},
  {Alexander}, {Chary}, {Starck}, {Brandt}, {Kitzbichler}, {MacDonald},
  {Nonino}, {Popesso}, {Stern}, \& {Vanzella}}]{elbaz07}
{Elbaz}, D., {Daddi}, E., {Le Borgne}, D., {et~al.} 2007, \aap, 468, 33

\bibitem[{{Faber} {et~al.}(2007){Faber}, {Willmer}, {Wolf}, {Koo}, {Weiner},
  {Newman}, {Im}, {Coil}, {Conroy}, {Cooper}, {Davis}, {Finkbeiner}, {Gerke},
  {Gebhardt}, {Groth}, {Guhathakurta}, {Harker}, {Kaiser}, {Kassin},
  {Kleinheinrich}, {Konidaris}, {Kron}, {Lin}, {Luppino}, {Madgwick},
  {Meisenheimer}, {Noeske}, {Phillips}, {Sarajedini}, {Schiavon}, {Simard},
  {Szalay}, {Vogt}, \& {Yan}}]{faber07}
{Faber}, S.~M., {Willmer}, C.~N.~A., {Wolf}, C., {et~al.} 2007, \apj, 665, 265

\bibitem[{{Furusawa} {et~al.}(2000){Furusawa}, {Shimasaku}, {Doi}, \&
  {Okamura}}]{furusawa00}
{Furusawa}, H., {Shimasaku}, K., {Doi}, M., \& {Okamura}, S. 2000, \apj, 534,
  624

\bibitem[{{Galametz} {et~al.}(2009){Galametz}, {Stern}, {Eisenhardt},
  {Brodwin}, {Brown}, {Dey}, {Gonzalez}, {Jannuzi}, {Moustakas}, \&
  {Stanford}}]{galametz09}
{Galametz}, A., {Stern}, D., {Eisenhardt}, P.~R.~M., {et~al.} 2009, \apj, 694,
  1309

\bibitem[{{Gebhardt} {et~al.}(2003){Gebhardt}, {Faber}, {Koo}, {Im}, {Simard},
  {Illingworth}, {Phillips}, {Sarajedini}, {Vogt}, {Weiner}, \&
  {Willmer}}]{gebhardt03}
{Gebhardt}, K., {Faber}, S.~M., {Koo}, D.~C., {et~al.} 2003, \apj, 597, 239

\bibitem[{{Gilbank} {et~al.}(2008){Gilbank}, {Yee}, {Ellingson}, {Gladders},
  {Loh}, {Barrientos}, \& {Barkhouse}}]{gilbank08}
{Gilbank}, D.~G., {Yee}, H.~K.~C., {Ellingson}, E., {et~al.} 2008, \apj, 673,
  742

\bibitem[{{Gobat} {et~al.}(2008){Gobat}, {Rosati}, {Strazzullo}, {Rettura},
  {Demarco}, \& {Nonino}}]{gobat08}
{Gobat}, R., {Rosati}, P., {Strazzullo}, V., {et~al.} 2008, \aap, 488, 853

\bibitem[{{Grazian} {et~al.}(2006){Grazian}, {Fontana}, {de Santis}, {Nonino},
  {Salimbeni}, {Giallongo}, {Cristiani}, {Gallozzi}, \& {Vanzella}}]{grazian06}
{Grazian}, A., {Fontana}, A., {de Santis}, C., {et~al.} 2006, \aap, 449, 951

\bibitem[{{Greve} {et~al.}(2007){Greve}, {Stern}, {Ivison}, {De Breuck},
  {Kov{\'a}cs}, \& {Bertoldi}}]{greve07}
{Greve}, T.~R., {Stern}, D., {Ivison}, R.~J., {et~al.} 2007, \mnras, 382, 48

\bibitem[{{Gunn} \& {Stryker}(1983)}]{gunn83}
{Gunn}, J.~E. \& {Stryker}, L.~L. 1983, \apjs, 52, 121

\bibitem[{{Hatch} {et~al.}(2009){Hatch}, {Overzier}, {Kurk}, {Miley},
  {R{\"o}ttgering}, \& {Zirm}}]{hatch09}
{Hatch}, N.~A., {Overzier}, R.~A., {Kurk}, J.~D., {et~al.} 2009, \mnras, 395,
  114

\bibitem[{{Hatch} {et~al.}(2008){Hatch}, {Overzier}, {R{\"o}ttgering}, {Kurk},
  \& {Miley}}]{hatch08}
{Hatch}, N.~A., {Overzier}, R.~A., {R{\"o}ttgering}, H.~J.~A., {Kurk}, J.~D.,
  \& {Miley}, G.~K. 2008, \mnras, 383, 931

\bibitem[{{Hopkins} {et~al.}(2003){Hopkins}, {Miller}, {Nichol}, {Connolly},
  {Bernardi}, {G{\'o}mez}, {Goto}, {Tremonti}, {Brinkmann}, {Ivezi{\'c}}, \&
  {Lamb}}]{hopkins03}
{Hopkins}, A.~M., {Miller}, C.~J., {Nichol}, R.~C., {et~al.} 2003, \apj, 599,
  971

\bibitem[{{Kennicutt}(1998)}]{kennicutt98}
{Kennicutt}, Jr., R.~C. 1998, \araa, 36, 189

\bibitem[{{Kodama} {et~al.}(2007){Kodama}, {Tanaka}, {Kajisawa}, {Kurk},
  {Venemans}, {De Breuck}, {Vernet}, \& {Lidman}}]{kodama07}
{Kodama}, T., {Tanaka}, I., {Kajisawa}, M., {et~al.} 2007, \mnras, 377, 1717

\bibitem[{{Koyama} {et~al.}(2007){Koyama}, {Kodama}, {Tanaka}, {Shimasaku}, \&
  {Okamura}}]{koyama07}
{Koyama}, Y., {Kodama}, T., {Tanaka}, M., {Shimasaku}, K., \& {Okamura}, S.
  2007, \mnras, 382, 1719

\bibitem[{{Kriek} {et~al.}(2008){Kriek}, {van Dokkum}, {Franx}, {Illingworth},
  {Marchesini}, {Quadri}, {Rudnick}, {Taylor}, {F{\"o}rster Schreiber},
  {Gawiser}, {Labb{\'e}}, {Lira}, \& {Wuyts}}]{kriek08}
{Kriek}, M., {van Dokkum}, P.~G., {Franx}, M., {et~al.} 2008, \apj, 677, 219

\bibitem[{{Kuntschner} {et~al.}(2002){Kuntschner}, {Smith}, {Colless},
  {Davies}, {Kaldare}, \& {Vazdekis}}]{kuntschner02}
{Kuntschner}, H., {Smith}, R.~J., {Colless}, M., {et~al.} 2002, \mnras, 337,
  172

\bibitem[{{Kurk} {et~al.}(2004{\natexlab{a}}){Kurk}, {Pentericci}, {Overzier},
  {R{\"o}ttgering}, \& {Miley}}]{kurk04b}
{Kurk}, J.~D., {Pentericci}, L., {Overzier}, R.~A., {R{\"o}ttgering}, H.~J.~A.,
  \& {Miley}, G.~K. 2004{\natexlab{a}}, \aap, 428, 817

\bibitem[{{Kurk} {et~al.}(2004{\natexlab{b}}){Kurk}, {Pentericci},
  {R{\"o}ttgering}, \& {Miley}}]{kurk04a}
{Kurk}, J.~D., {Pentericci}, L., {R{\"o}ttgering}, H.~J.~A., \& {Miley}, G.~K.
  2004{\natexlab{b}}, \aap, 428, 793

\bibitem[{{Kurk} {et~al.}(2000){Kurk}, {R{\"o}ttgering}, {Pentericci}, {Miley},
  {van Breugel}, {Carilli}, {Ford}, {Heckman}, {McCarthy}, \&
  {Moorwood}}]{kurk00}
{Kurk}, J.~D., {R{\"o}ttgering}, H.~J.~A., {Pentericci}, L., {et~al.} 2000,
  \aap, 358, L1

\bibitem[{{Lehmer} {et~al.}(2005){Lehmer}, {Brandt}, {Alexander}, {Bauer},
  {Schneider}, {Tozzi}, {Bergeron}, {Garmire}, {Giacconi}, {Gilli}, {Hasinger},
  {Hornschemeier}, {Koekemoer}, {Mainieri}, {Miyaji}, {Nonino}, {Rosati},
  {Silverman}, {Szokoly}, \& {Vignali}}]{lehmer05}
{Lehmer}, B.~D., {Brandt}, W.~N., {Alexander}, D.~M., {et~al.} 2005, \apjs,
  161, 21

\bibitem[{{Lidman} {et~al.}(2008){Lidman}, {Rosati}, {Tanaka}, {Strazzullo},
  {Demarco}, {Mullis}, {Ageorges}, {Kissler-Patig}, {Petr-Gotzens}, \&
  {Selman}}]{lidman08}
{Lidman}, C., {Rosati}, P., {Tanaka}, M., {et~al.} 2008, \aap, 489, 981

\bibitem[{{Madau}(1995)}]{madau95}
{Madau}, P. 1995, \apj, 441, 18

\bibitem[{{Mei} {et~al.}(2009){Mei}, {Holden}, {Blakeslee}, {Ford}, {Franx},
  {Homeier}, {Illingworth}, {Jee}, {Overzier}, {Postman}, {Rosati}, {Van der
  Wel}, \& {Bartlett}}]{mei09}
{Mei}, S., {Holden}, B.~P., {Blakeslee}, J.~P., {et~al.} 2009, \apj, 690, 42

\bibitem[{{Miley} \& {De Breuck}(2008)}]{miley08}
{Miley}, G. \& {De Breuck}, C. 2008, \aapr, 15, 67

\bibitem[{{Miley} {et~al.}(2006){Miley}, {Overzier}, {Zirm}, {Ford}, {Kurk},
  {Pentericci}, {Blakeslee}, {Franx}, {Illingworth}, {Postman}, {Rosati},
  {R{\"o}ttgering}, {Venemans}, \& {Helder}}]{miley06}
{Miley}, G.~K., {Overzier}, R.~A., {Zirm}, A.~W., {et~al.} 2006, \apjl, 650,
  L29

\bibitem[{{Nakata} {et~al.}(2005){Nakata}, {Kodama}, {Shimasaku}, {Doi},
  {Furusawa}, {Hamabe}, {Kimura}, {Komiyama}, {Miyazaki}, {Okamura}, {Ouchi},
  {Sekiguchi}, {Ueda}, {Yagi}, \& {Yasuda}}]{nakata05}
{Nakata}, F., {Kodama}, T., {Shimasaku}, K., {et~al.} 2005, \mnras, 357, 1357

\bibitem[{{Pentericci} {et~al.}(2002){Pentericci}, {Fan}, {Rix}, {Strauss},
  {Narayanan}, {Richards}, {Schneider}, {Krolik}, {Heckman}, {Brinkmann},
  {Lamb}, \& {Szokoly}}]{pentericci02}
{Pentericci}, L., {Fan}, X., {Rix}, H., {et~al.} 2002, \aj, 123, 2151

\bibitem[{{Pentericci} {et~al.}(2000){Pentericci}, {Kurk}, {R{\"o}ttgering},
  {Miley}, {van Breugel}, {Carilli}, {Ford}, {Heckman}, {McCarthy}, \&
  {Moorwood}}]{pentericci00}
{Pentericci}, L., {Kurk}, J.~D., {R{\"o}ttgering}, H.~J.~A., {et~al.} 2000,
  \aap, 361, L25

\bibitem[{{Pentericci} {et~al.}(1997){Pentericci}, {Roettgering}, {Miley},
  {Carilli}, \& {McCarthy}}]{pentericci97}
{Pentericci}, L., {Roettgering}, H.~J.~A., {Miley}, G.~K., {Carilli}, C.~L., \&
  {McCarthy}, P. 1997, \aap, 326, 580

\bibitem[{{Pentericci} {et~al.}(1998){Pentericci}, {Roettgering}, {Miley},
  {Spinrad}, {McCarthy}, {van Breugel}, \& {Macchetto}}]{pentericci98}
{Pentericci}, L., {Roettgering}, H.~J.~A., {Miley}, G.~K., {et~al.} 1998, \apj,
  504, 139

\bibitem[{{Polletta} {et~al.}(2007){Polletta}, {Tajer}, {Maraschi},
  {Trinchieri}, {Lonsdale}, {Chiappetti}, {Andreon}, {Pierre}, {Le F{\`e}vre},
  {Zamorani}, {Maccagni}, {Garcet}, {Surdej}, {Franceschini}, {Alloin},
  {Shupe}, {Surace}, {Fang}, {Rowan-Robinson}, {Smith}, \&
  {Tresse}}]{polletta07}
{Polletta}, M., {Tajer}, M., {Maraschi}, L., {et~al.} 2007, \apj, 663, 81

\bibitem[{{Postman} {et~al.}(2005){Postman}, {Franx}, {Cross}, {Holden},
  {Ford}, {Illingworth}, {Goto}, {Demarco}, {Rosati}, {Blakeslee}, {Tran},
  {Ben{\'{\i}}tez}, {Clampin}, {Hartig}, {Homeier}, {Ardila}, {Bartko},
  {Bouwens}, {Bradley}, {Broadhurst}, {Brown}, {Burrows}, {Cheng}, {Feldman},
  {Golimowski}, {Gronwall}, {Infante}, {Kimble}, {Krist}, {Lesser}, {Martel},
  {Mei}, {Menanteau}, {Meurer}, {Miley}, {Motta}, {Sirianni}, {Sparks}, {Tran},
  {Tsvetanov}, {White}, \& {Zheng}}]{postman05}
{Postman}, M., {Franx}, M., {Cross}, N.~J.~G., {et~al.} 2005, \apj, 623, 721

\bibitem[{{Press} \& {Schechter}(1974)}]{press74}
{Press}, W.~H. \& {Schechter}, P. 1974, \apj, 187, 425

\bibitem[{{Reddy} {et~al.}(2006){Reddy}, {Steidel}, {Fadda}, {Yan}, {Pettini},
  {Shapley}, {Erb}, \& {Adelberger}}]{reddy06}
{Reddy}, N.~A., {Steidel}, C.~C., {Fadda}, D., {et~al.} 2006, \apj, 644, 792

\bibitem[{{Salpeter}(1955)}]{salpeter55}
{Salpeter}, E.~E. 1955, \apj, 121, 161

\bibitem[{{Santini} {et~al.}(2009){Santini}, {Fontana}, {Grazian}, {Salimbeni},
  {Fiore}, {Fontanot}, {Boutsia}, {Castellano}, {Cristiani}, {De Santis},
  {Gallozzi}, {Giallongo}, {Menci}, {Nonino}, {Paris}, {Pentericci}, \&
  {Vanzella}}]{santini09}
{Santini}, P., {Fontana}, A., {Grazian}, A., {et~al.} 2009, ArXiv e-prints

\bibitem[{{Schlegel} {et~al.}(1998){Schlegel}, {Finkbeiner}, \&
  {Davis}}]{schlegel98}
{Schlegel}, D.~J., {Finkbeiner}, D.~P., \& {Davis}, M. 1998, \apj, 500, 525

\bibitem[{{Seymour} {et~al.}(2007){Seymour}, {Stern}, {De Breuck}, {Vernet},
  {Rettura}, {Dickinson}, {Dey}, {Eisenhardt}, {Fosbury}, {Lacy}, {McCarthy},
  {Miley}, {Rocca-Volmerange}, {R{\"o}ttgering}, {Stanford}, {Teplitz}, {van
  Breugel}, \& {Zirm}}]{seymour07}
{Seymour}, N., {Stern}, D., {De Breuck}, C., {et~al.} 2007, \apjs, 171, 353

\bibitem[{{Sirianni} {et~al.}(2005){Sirianni}, {Jee}, {Ben{\'{\i}}tez},
  {Blakeslee}, {Martel}, {Meurer}, {Clampin}, {De Marchi}, {Ford}, {Gilliland},
  {Hartig}, {Illingworth}, {Mack}, \& {McCann}}]{sirianni05}
{Sirianni}, M., {Jee}, M.~J., {Ben{\'{\i}}tez}, N., {et~al.} 2005, \pasp, 117,
  1049

\bibitem[{{Springel} {et~al.}(2005){Springel}, {White}, {Jenkins}, {Frenk},
  {Yoshida}, {Gao}, {Navarro}, {Thacker}, {Croton}, {Helly}, {Peacock}, {Cole},
  {Thomas}, {Couchman}, {Evrard}, {Colberg}, \& {Pearce}}]{springel05}
{Springel}, V., {White}, S.~D.~M., {Jenkins}, A., {et~al.} 2005, \nat, 435, 629

\bibitem[{{Stanford} {et~al.}(2006){Stanford}, {Romer}, {Sabirli}, {Davidson},
  {Hilton}, {Viana}, {Collins}, {Kay}, {Liddle}, {Mann}, {Miller}, {Nichol},
  {West}, {Conselice}, {Spinrad}, {Stern}, \& {Bundy}}]{stanford06}
{Stanford}, S.~A., {Romer}, A.~K., {Sabirli}, K., {et~al.} 2006, \apjl, 646,
  L13

\bibitem[{{Steidel} {et~al.}(2005){Steidel}, {Adelberger}, {Shapley}, {Erb},
  {Reddy}, \& {Pettini}}]{steidel05}
{Steidel}, C.~C., {Adelberger}, K.~L., {Shapley}, A.~E., {et~al.} 2005, \apj,
  626, 44

\bibitem[{{Tanaka} {et~al.}(2008){Tanaka}, {Finoguenov}, {Kodama}, {Morokuma},
  {Rosati}, {Stanford}, {Eisenhardt}, {Holden}, \& {Mei}}]{tanaka08}
{Tanaka}, M., {Finoguenov}, A., {Kodama}, T., {et~al.} 2008, \aap, 489, 571

\bibitem[{{Tanaka} {et~al.}(2005){Tanaka}, {Kodama}, {Arimoto}, {Okamura},
  {Umetsu}, {Shimasaku}, {Tanaka}, \& {Yamada}}]{tanaka05}
{Tanaka}, M., {Kodama}, T., {Arimoto}, N., {et~al.} 2005, \mnras, 362, 268

\bibitem[{{Tanaka} {et~al.}(2007){Tanaka}, {Kodama}, {Kajisawa}, {Bower},
  {Demarco}, {Finoguenov}, {Lidman}, \& {Rosati}}]{tanaka07a}
{Tanaka}, M., {Kodama}, T., {Kajisawa}, M., {et~al.} 2007, \mnras, 377, 1206

\bibitem[{{Thomas} {et~al.}(2005){Thomas}, {Maraston}, {Bender}, \& {Mendes de
  Oliveira}}]{thomas05}
{Thomas}, D., {Maraston}, C., {Bender}, R., \& {Mendes de Oliveira}, C. 2005,
  \apj, 621, 673

\bibitem[{{Venemans} {et~al.}(2007){Venemans}, {R{\"o}ttgering}, {Miley}, {van
  Breugel}, {de Breuck}, {Kurk}, {Pentericci}, {Stanford}, {Overzier}, {Croft},
  \& {Ford}}]{venemans07}
{Venemans}, B.~P., {R{\"o}ttgering}, H.~J.~A., {Miley}, G.~K., {et~al.} 2007,
  \aap, 461, 823

\bibitem[{{Zirm} {et~al.}(2008){Zirm}, {Stanford}, {Postman}, {Overzier},
  {Blakeslee}, {Rosati}, {Kurk}, {Pentericci}, {Venemans}, {Miley},
  {R{\"o}ttgering}, {Franx}, {van der Wel}, {Demarco}, \& {van
  Breugel}}]{zirm08}
{Zirm}, A.~W., {Stanford}, S.~A., {Postman}, M., {et~al.} 2008, \apj, 680, 224

\end{thebibliography}

%-------------------------------------------------------
%% \appendix
%% \section{A revised analysis of the two H$\alpha$ objects in PKS1138 reported in \citet{doherty09}}

\end{document}